%% file: main_v3.tex
\documentclass[aps,reprint]{revtex4-2}
\usepackage{amsmath}
\usepackage{amsfonts}
\usepackage{amssymb}
\usepackage{graphicx}
\usepackage[utf8]{inputenc}
\usepackage{enumitem}
\usepackage{color}
\usepackage[percent]{overpic}
\usepackage{bm}
\usepackage{rotating}
\usepackage{pbox}
\usepackage{array}
\usepackage[normalem]{ulem} 
\usepackage{physics}
\usepackage{xr-hyper}
\usepackage{hyperref}
\usepackage{xcolor}

\begin{document}

\title{Resonant fragility and nonresonant robustness of Floquet eigenstates in kicked spin systems}
\author{Jes\'us A. Segura-Landa \textsuperscript{1,2}}
\author{Meenu Kumari\textsuperscript{3,4}}
\email{mkumari@uwaterloo.ca}
\author{Daniel J. Nader\textsuperscript{5}}
\author{Sercan H\"usn\"ugil\textsuperscript{4,6}}
\author{Ali SaraerToosi\textsuperscript{4,7}}
\author{Sergio Lerma-Hern\'andez\textsuperscript{1}}
\email{slerma@uv.mx}
 \affiliation{\textsuperscript{1}Facultad  de F\'isica, Universidad Veracruzana,  Campus Arco Sur, Paseo 112, C.P. 91097  Xalapa, Mexico.}
 \affiliation{\textsuperscript{2}Instituto de Ciencias Nucleares, Universidad Nacional Aut\'onoma de M\'exico, Apdo. Postal 70-543, C.P. 04510  Cd. Mx., Mexico}
\affiliation{\textsuperscript{3}Digital Technologies, National Research Council Canada}
\affiliation{\textsuperscript{4}Perimeter Institute for Theoretical Physics, Waterloo ON N2L 2Y5, Canada}
\affiliation{\textsuperscript{5}Department of Optics, Faculty of Science, Palacky University, Olomouc, 77900, Czech Republic}
\affiliation{\textsuperscript{6}Department of Physics and Astronomy, University of Waterloo, Waterloo, Ontario, N2L 3G1, Canada}
\affiliation{\textsuperscript{7}Department of Computer Science, University of Toronto, 40 St. George St., Toronto, ON, M5S 2E4, Canada}

\begin{abstract}
In classical systems, the Kolmogorov–Arnold–Moser (KAM) theorem  establishes that resonant tori of integrable Hamiltonians are destroyed by any nonintegrable perturbation, whereas nonresonant tori are only deformed up to a finite value of the perturbation parameter. In this contribution, we identify a quantum analog of this differentiated sensitivity for one-degree-of-freedom spin Hamiltonians subject to periodic instantaneous kicks. After detecting quantum signatures of resonances in the participation ratio and in the quasiprobability phase-space distribution of Floquet eigenstates of the perturbed Hamiltonian, we show that eigenstates of the unperturbed Hamiltonian exhibit greater sensitivity against the perturbation when they satisfy a resonant condition. The sensitivity is quantified through the  fidelity between perturbed and unperturbed eigenstates. This differentiated sensitivity becomes increasingly pronounced as the system size grows. Our findings are supported by numerical results and insights from analytical calculations  based on unitary perturbation theory. Although our analysis focuses on kicked models, the mechanism could be extended to more general periodic drivings, providing a preliminary step toward a quantum counterpart of the classical breaking of resonant tori.
\end{abstract}
\maketitle


\section {Introduction} 
A system is Liouville integrable in classical physics if it has as many independent conserved quantities as degrees of freedom, and their Poisson brackets vanish pairwise \cite{mathematicalmethods}. 
The Kolmogorov-Arnold-Moser (KAM) theorem forms a cornerstone in classical physics demonstrating that weakly perturbed integrable systems retain most of the properties of integrability for sufficiently small perturbations \cite{mathematicalmethods,Livi2003,Weissert1997,Sreeram2024Review}. In quantum physics, while the extremes of fully integrable and globally chaotic systems are relatively well understood, the integrability-to-chaos transition and the dynamics of weakly perturbed integrable systems remain elusive. The search for a quantum analog of the KAM theorem in few and many-body systems has been ongoing for decades \cite{Hose1983,Reichl1987,ReichlLin88,BCK2015,RudolfPRL2017,SerbynPRX2020,BFH2021,GopalakrishnanPRB2022,YHDHO2022,YYJWZ2022}. Recent studies have explored weakly perturbed integrable quantum systems from various perspectives, including the effects of perturbation on the conserved quantities \cite{BCK2015,BFH2021,KurlovPRB2022,KurlovPRB2023}, eigenvalues and spectral statistics \cite{ReichlLin88,RudolfPRL2017,YHDHO2022}, and eigenstates \cite{SelsPRX2020}. Additionally, dynamical probes such as out-of-time-order correlators (OTOCs) \cite{RudolfPRL2017},  
relaxation to thermal states  \cite{RobinsonPRB2016,SerbynPRX2020,BenjaminPRX2018} and hydrodynamics\cite{VasseurPRB2020,Vasseur2021} have also been studied in such systems. These studies have revealed phenomena such as prethermalization plateaus, where local observables relax to nonthermal values at intermediate times \cite{RobinsonPRB2016}, scaling properties of perturbation strengths as a function of system size when the system transitions to chaos \cite{GopalakrishnanPRB2022}, and the persistence of quasiconserved charges up to specific time scales \cite{KurlovPRB2022}. These findings underscore the intricate nature of the integrability-chaos transition \cite{YurovskyPRL2023}, offering insights into how classical KAM-like stability might manifest in quantum systems. 

In this work, we investigate the weak breaking of integrability in a class of many-body systems subjected to periodic kicks. Owing to their SU(2) symmetry, these systems can be reduced to models with only a few degrees of freedom, which greatly facilitates both numerical and analytical treatment and allows the exploration of very large system sizes. We first identify the quantum analog of the classical resonant condition as those cases where the driving period is an integer multiple of the inverse transition frequency between two eigenstates of the unperturbed system. Our analysis reveals that eigenstates that closely approximate a resonant condition are significantly more fragile under perturbation, rapidly delocalizing in the unperturbed eigenbasis as the perturbation is introduced. In contrast, states far from resonance remain strongly localized near a single unperturbed basis state. To quantify this differentiated sensitivity, we employ and examine the maximal overlap, or fidelity, between the perturbed (Floquet) eigenstates and the unperturbed basis states. Specifically, we analyze how the perturbation strength must change (decrease) as the system size increases to maintain a fixed fidelity value of $1/2$. Our results reveal a significant difference in this sensitivity between states close to a resonance and those further away, which becomes more pronounced with the increase of system size. Using unitary perturbation theory, we show that the underlying mechanism of rapid delocalization for eigenstates close to resonance is the emergence of small denominators. Our study successfully extends the insights of the KAM theorem into the quantum realm by enabling the classification of eigenstates of a quantum integrable system into two distinct classes based on their sensitivity to perturbations. This classification represent a preliminary step toward establishing a quantum counterpart of the KAM theorem. 

This work is organized as follows. Section \ref{Sec:kicked models} introduces   the kicked models under study. Section \ref{sec:QuantumResonances} discusses the quantum analog of classical resonant conditions and examine some of their manifestations. In section \ref{Sec:epmax}, we define the parameter used to quantify the sensitivity of   states to  perturbations, analyze  its dependence on  system size,  and show how unitary perturbation theory  explains the numerical results.  Finally,  Section \ref{Sec:Discussion} presents our conclusions.

\section{Kicked models}
\label{Sec:kicked models}
Specifically, we study time-dependent spin-$J$ Hamiltonians of the following form: 
\begin{equation}
\label{Ham}
\hat{H}(t) = \hat{H}_0(\vec{J})+\epsilon\hat{K}(\vec{J})\sum_{n=-\infty}^{\infty}\delta(t-n\tau),
\end{equation}
where 
$\hat{J}_{x,y,z}=\frac{1}{2}\sum_{n=1}^{N} \hat{\sigma}_{x,y,z}^{(n)}$, with $\sigma_i^{(n)}$  Pauli operators, $[\hat{J}_i,\hat{J}_j]=i\epsilon_{ijk}\hat{J}_k$  $(\hbar=1)$, and $J(J+1)$ is the eigenvalue of $\hat{J}_x^2+\hat{J}_y^2+\hat{J}_z^2$. We consider the  subspace with largest $J$ value, $J=N/2$. $\hat{H}_0(\vec{J})$ and $\hat{K}(\vec{J})$ describe quantum systems with well-defined classical limits ($1/\hbar_\text{eff}\equiv J \rightarrow\infty$) with one-degree-of-freedom. 
$\tau$ and $\epsilon$ represent the time period and strength of the kick, respectively.
Such systems have been widely studied in contexts including quantum integrability and chaos \cite{Haake,Manuel2021PRE,Sinha2016, meenu2,Dogra2019PRE,Santhanam2018PRE,RobnikPRE2023,Anand2024PRR,Bhosale2024PRB,Sinha2024Review}, entanglement \cite{entanglement2004PRE,Ghose2008PRA,KumariThesis,Arjendu2017PRE,Silvia2020PRA,Bhosale2024entanglement}, quantum simulation and trotterization \cite{sieberer2019digital,Chinni2022,ManuelPRL2020},  time crystals \cite{KlmgCrystalPRB2017,Poggi2022PRR}, and have also been experimentally realized \cite{Chaudhury2009nature,Neill2016,Krithika2019PRE}. Classically, the periodic kicks introduce non-integrability, modulated by the dimensionless perturbation strength $\epsilon$ \cite{Haake}. For small \(\epsilon\), the system remains quasi-integrable, preserving non-resonant tori with  slight deformations as predicted by the classical KAM theorem. On the other hand, resonant tori are destroyed, leading to the emergence of elliptic and hyperbolic fixed points according to the Poincaré-Birkhoff theorem \cite{birkhoff1926extension}. In the quantum counterpart, manifestations of such behavior are expected to occur \cite{wisniacki,billam2009quantum,Schlagheck2015}, particularly on the phase-space representation of the quantized stationary states of the complete Hamiltonian. Using Floquet theory  
\cite{floquet1883equations,GRIFONI1998229}, we classify the quantum system's eigenstates into resonant and non-resonant via numerical and analytical calculations using unitary perturbation theory \cite{Peres,Chinni2022}. The stroboscopic stationary states are identified as eigenvectors, $\ket{f_k}$, of the Floquet operator, $\hat{F}\ket{f_k}=\exp(-i\phi_k
)\ket{f_k}$, with quasi-energies \(\phi_k\) and
\begin{equation} \hat{F} = \hat{F}_o \hat{F}_k = \exp\left(-i\tau\hat{H}_0\right)\exp\left(-i\epsilon\hat{K}\right). \end{equation}
Our findings reveal distinct sensitivity of eigenstates of $\hat{H_0}$ to time-periodic perturbations, characterized by the strength of the periodic kicks \(\epsilon\) relative to the system size \(J\). For sufficiently large J values, we find that resonant eigenstates  begin to significantly  spread over the unperturbed $\hat{H}_0$ eigenbasis at a perturbation strength  that vanishes as $\epsilon\propto 1/J^2$, whereas  for  non-resonant states this threshold scales only  as $\epsilon\propto 1/J$. Such a classification of Floquet eigenstates is consistent with recent studies on quantum many-body resonances \cite{PolkovnikovPRB2016, CauxPRL2018, ChalkerPRB2021, GopalakrishnanPRB2022, HusePRB2022}, and we uniquely quantify their sensitivity relative to system size.  

\section{Quantum resonances and their manifestations}
\label{sec:QuantumResonances}
\begin{figure}[t]
      \includegraphics[width=\linewidth]{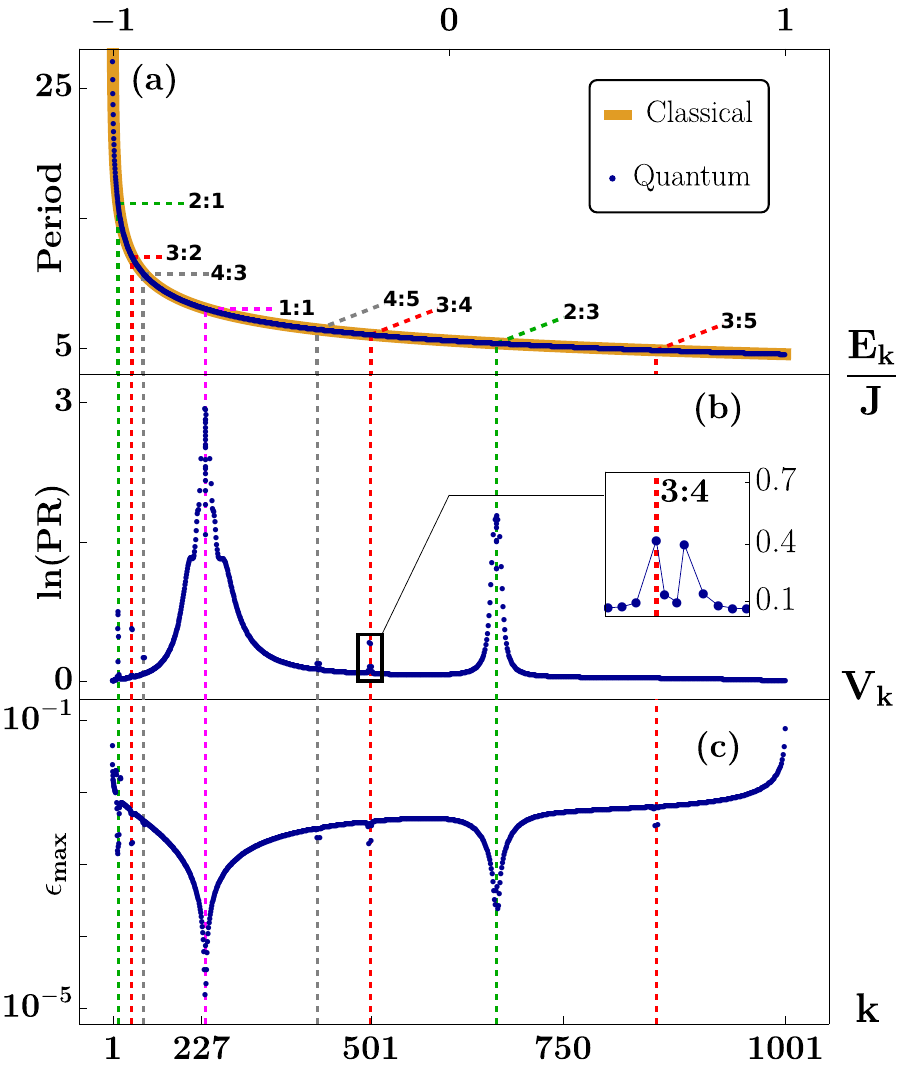}
\caption{(a) Classical time period (orange curve) and quantum periods, $T_{k,1}=2\pi/(E_{k+1}-E_k)$ (blue dots), as a function of energy of $h_0$ (classical) and scaled energies, $\bar{E}_k/J = (E_{k+1} +  E_k)/(2J) $ (quantum).  (b) Logarithm of the participation ratio of the Floquet eigenstates, $\ket{f_k}$, with respect to the $\hat{H}_0$ eigenbasis plotted as a function of $V_k=\langle f_k|\hat{H}_0| f_k\rangle/J$. The inset is a zoom  of the $3{:}4$ resonance, showing that third neighbours states are involved in this resonance. (c) Maximum perturbation strength $\epsilon_{k,\text{max}}$, Eq.\eqref{eq: max epsilon}, as a function of the index $k$ of the Floquet eigenstates $|f_k\rangle$ associated, respectively, to the states $|E_k\rangle$.
Dashed lines with labels $m{:}n$ in the panels  indicate some rational multiples  of the  kick period $m\tau/n$ and the energies associated, $E_{m:n}$. Parameters used: $J=500$ and  $\tau=8$ for (a-c), and $\epsilon=10^{-3}$ for (b). Data for (c) available at \cite{Raw_data_doi}}
 \label{fig:1}
\end{figure}
\begin{figure}[t]    
   \includegraphics[width=\linewidth]{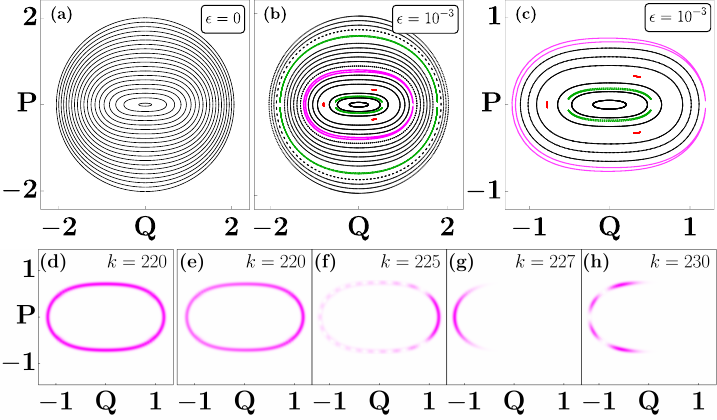} 
   \caption{\textit{First row}: stroboscopic Poincar\'e sections of the classical LMG Hamiltonian (a) without and (b) with a small perturbation; panel (c) is a zoom of the low-energy region of (b). \textit{Second row}: (d) Husimi function of a representative  LMG eigenstate in that energy region, and (e)-(h) of Floquet eigenstates close to 1:1 resonance. Parameters used: $\tau=8$ and  $\epsilon=10^{-3}$, and for second row $J=500$.} 
  \label{fig:2}
\end{figure}
Classically, in a one-degree-of-freedom periodically kicked system with period $\tau$, and  non-linear Hamilton's equations of motion, an $m{:}n$ resonance  takes place when the resonant condition is fulfilled
\begin{equation}\label{eq: resonant condition}
    nT(E) = m\tau, \quad n,m\in\mathbb{Z},
\end{equation}
where $T(E)$ is the classical time period of the orbit of $H_0$ with energy $E$. Such an $m{:}n$ resonance is associated with an $m$-cycle in the phase space. In quantum mechanics, it is also possible to establish a resonant condition. For this goal,  we first define  the quantum periods  $T_{k,m}$ as  ($m>0$)
\begin{equation}\label{eq: quantum periods}
T_{k,m} = \frac{2\pi}{E_{k+m} - E_{k}}\xrightarrow[]{J\to \infty}\frac{2\pi}{m(E_{k+1}-E_k)}=\frac{T_{k,1}}{m},
\end{equation}
where $E_k$ are the ordered quantized energy levels of Hamiltonian $\hat{H}_0$, i.e. $\hat{H}_0\ket{E_k}=E_k\ket{E_k}$ with $E_{k+1}>E_k$. This quantum period is the inverse of the transition frequency between states with indices $k$ and $k+m$. For $m=1$ (nearest neighbors), the equation above reduces to the semiclassical approximation of the  classical  periods \cite{landau1991quantum,NADER_Roberto} $T_{k,1}\approx T(\bar{E})$ with $\bar{E}=(E_{k+1}+E_k)/2$. Moreover, from Eq.~(\ref{eq: quantum periods}), it follows that the quantum resonant condition is
\begin{equation}
    n T_{k,m}=\tau \text{\ \ \ or\ \ } \tau(E_{k+m}-E_{k})=2\pi n.  
     \label{eq:QR}
\end{equation}
The integer number $m$ enters into the previous quantum resonant condition by involving $m$-th neighbor states [
see inset in Fig.\ref{fig:1}(b)]. For systems with  one-degree-of-freedom in the classical  limit ($J\rightarrow  \infty$), the energy difference satisfies    $\lim_{J\rightarrow\infty}(E_{k+m}-E_{k})\approx m (E_{k+1}-E_k)$, and thus   Eq.\eqref{eq:QR} reduces to the classical resonance condition~\eqref{eq: resonant condition}. 

In order to illustrate the quantum and classical manifestations of the resonances in an experimentally realizable scenario, we consider as the time-independent term, the Hamiltonian of the Lipkin-Meshkov-Glick (LMG) model \cite{LIPKIN1} in the pseudo spin representation \cite{NaderAvoided,ESQPT,Ribeiro2008,segura2024quantum} 
\begin{equation}
    \hat{H}_0=\hat{J}_z+\frac{\gamma_x}{2 J-1}\hat{J}_x^2  + \frac{\gamma_y}{2J-1}\hat{J}_y^2,
\end{equation}
with control parameters $\gamma_x\!\!=\!\!-0.95$ and $\gamma_y\!\!=\!\!0$, and the kicking term in \eqref{Ham} as $\hat{K}\!\!=\!\!\hat{J}_z+\hat{J}_x$. This selection is beneficial for the manifestations of the resonances for two reasons: i) the density of states, $\rho(\bar{E})= T_{k,1}/(2\pi)$ \cite{NADER_Roberto}, of the LMG model for these parameters is free of singularities and therefore the so-called Excited State Quantum Phase Transitions (ESQPTs) \cite{cejnar2021excited} are absent, and ii) the chosen kick operator has non zero perturbation matrix elements $\langle E_{k+m}|\hat{K}|E_k\rangle\not=0$ whose modulus decreases with $m$, which turns out to be essential towards a clearer
manifestation of the resonances as we will show  in Subsection IV.~B.  More general scenarios in which the spectrum of $\hat {H}_0$ exhibits degeneracies or avoided crossings, together with generic perturbation operators, lie beyond the scope of this study and will be addressed in future work.  

The Hamiltonian in the classical limit can be obtained straightforwardly by taking the expectation  value of the Hamiltonian  in  Bloch coherent states \cite{Ribeiro2006,Ribeiro2008,bastarrachea2014comparative}
\begin{equation}
    \begin{split}
        h(Q,P,t)&\equiv\lim_{J\rightarrow\infty}\frac{\langle\alpha |\hat{H}(t)|\alpha \rangle}{J}\\
        &=h_0(Q,P)+\epsilon k(Q,P)\sum\delta(t-n\tau)
    \end{split}
\end{equation}
where
$$
\ket{\alpha}=\frac{e^{\alpha\hat{J}_+}}{(1+\abs{\alpha}^2)^J}\ket{J,\,-J},
$$
and the  complex parameter $\alpha$ is expressed in terms of classical canonical variables $Q$ and $P$ as $$\alpha(Q,P)=\frac{(Q-i P)}{\sqrt{4-(Q^2+P^2)}},$$ 
which are constrained by $Q^2+P^2\leq 2^2.$

 The classical unperturbed hamiltonian  and kicking term are, respectively, 
\begin{align}
    h_0&=\frac{Q^2+P^2}{2}-1+\left(1-\frac{Q^2+P^2}{4}\right)\left(\frac{\gamma_x Q^2+\gamma_y P^2}{2}\right),&\nonumber\\
    k&=\left(\frac{Q^2+P^2}{2}-1\right)+\left(Q\sqrt{1-\frac{Q^2+P^2}{4}}\right).&
\end{align}

In Fig.\ref{fig:1}(a), we show the time period $T$ of the classical LMG Hamiltonian $h_0$  (see Sec. I in \cite{SM}) along with the quantum periods $T_{k,1}=2 \pi/(E_{k+1}-E_k)$ plotted versus the respective mean energy $\bar{E}_k/J=(E_{k+1}+E_k)/(2J)$. Quantum and classical periods are barely distinguishable. In the same figure, horizontal and diagonal  dashed lines indicate some rational multiples of the kick period $m\tau/n$ with $\tau=8$.  Resonances appear for trajectories of $h_0$ with energy $E_{m:n}/J$ (corresponding vertical dashed lines). Consequently, the corresponding tori are destroyed as  it is illustrated in the stroboscopic Poincaré sections shown in panels (a)-(c) of Fig.~\ref{fig:2}, where resonant trajectories are highlighted in magenta for resonance 1:1; in green for resonances 2:1 and  2:3;  and in red for resonance 3:2. 

Quantum resonances can be revealed by studying the localization of the Floquet eigenstates on the $\hat{H}_0$ energy eigenbasis $\ket{E_k}$ through the Participation Ratio \cite{PR,Gonzalez2025}, defined as \begin{equation}\text{PR}_k=\frac{1}{\sum_{n}\abs{\braket{E_n}{f_k}}^4}.\end{equation}
This quantity is plotted as a function of the expectation value $V_k=\mel{f_k}{\hat{H}_0}{f_k}/J$, in Fig. \ref{fig:1}(b). The resonances manifest as a sudden increase in $\text{PR}_k$, which reflects that, close to a resonant condition, Floquet eigenstates spread across the $\hat{H}_0$ eigenbasis. Quantum resonances manifest even if, due to the discreteness of the values of \(T_{k,m}\), the quantum resonant condition may not be exactly fulfilled. Meanwhile, in the classical picture, the continuity of \(T
\) guarantees that the resonant conditions are always exactly satisfied. When the quantum resonant condition (\ref{eq:QR}) is exactly satisfied, the resonant effects are more prominent~(see Sec.II in  \cite{SM}). 

Quantum manifestations of the resonances can also be observed in the phase space quasi-probability distribution functions of the Floquet eigenstates, such as Husimi functions. The Husimi function, $\mathcal{Q}_k(P,Q)= \abs{\braket{\alpha(Q,P)}{f_k}}^2,$ of Floquet eigenstates close to the 1:1 resonance are shown in    Figs.~\ref{fig:2}(e)-(h). Eigenstate with index \(k = 220\) is far enough from the resonant condition, such that its Husimi function is very similar to those of the LMG eigenstates, $\tilde{\mathcal{Q}}_n(P,Q)= |\langle \alpha(Q,P)|E_n\rangle|^2$,  as the one in Fig.\ref{fig:2}(d). In contrast, states closer to the resonance (\(k = 225, 227,\) and \(230\)) concentrate within the region of the classical broken tori, exhibiting a pattern of nodes and antinodes, as previously reported in Refs.~\cite{Reichl2021,wisniacki,billam2009quantum, Schlagheck2015} for different models. 

\section{Gauging the fragility of eigenstates to the perturbation}
\label{Sec:epmax}
Classically,  the KAM theorem states that a resonant condition destroys the $h_0$ tori even under  an infinitesimal perturbation $\epsilon$. 
In contrast, in the quantum regime,  Floquet eigenstates at infinitesimal $\epsilon$ can be unequivocally identified with the  energy eigenstates of the unperturbed Hamiltonian $\hat{H}_0$. As $\epsilon$  increases, this identification for $\ket{f_k}$ can be guaranteed   if   its maximum overlap with the eigenstates of  $\hat{H}_0$  satisfies \cite{Hose1983} 
$$\mathcal{F}_{k,\text{max}}\equiv\max\limits_{n}\abs{\braket{f_k}{E_{n}}}^2>\frac{1}{2}.$$
The smallest perturbation strength for which the  equality
\begin{equation}
\mathcal{F}_{k,\text{max}}=1/2
\label{eq: max epsilon}    
\end{equation}
holds defines a threshold   $\epsilon$   below  which the Floquet eigenstate   $|f_k\rangle$ can   be safely  associated with a single eigenstate of $\hat{H_0}$.  We denote this  value by  $\epsilon_{k,\text{max}}$,  which quantifies  the degree of fragility of the $\hat{H_0}$ eigenstates against  the perturbation. It should be  noted  that although $\epsilon_{k,\text{max}}$ defines a perturbation interval, $(0,\epsilon_{k,\text{max}})$, within which   Floquet eigenstates can be reliably associated with a single eigenstate of $\hat{H}_0$, this identification may in some cases extend   beyond $\epsilon_{k,\text{max}}$. For resonant states, however,  the condition $\mathcal{F}_{k,\text{max}}=1/2$ also marks the value of $\epsilon$ at which  the Floquet eigenstate spreads equally over  the two eigenstates of  $\hat{H}_0$ involved in the resonance (see next section), thereby  breaking the one-to-one correspondence. In contrast, for non-resonant eigenstates this breakdown does not occur, and the  one-to-one identification can persist well beyond $\epsilon_{k,\text{max}}$.  
  
The threshold $\epsilon_{k,\text{max}}$  is significantly smaller for states close to a quantum resonant condition. This is illustrated in Fig. \ref{fig:1}(c), where $\epsilon_{k,\text{max}}$ is plotted for all the Floquet eigenstates against the energy of the corresponding $\hat{H}_0$ eigenstate with which they are associated. From this figure,  it is clear that $\epsilon_{k,\text{max}}$ is highly sensitive to the quantum resonant condition, exhibiting a sudden decrease close to the resonances. These dips become more prominent as we increase $J$ and approach the classical limit.  In  the following section, we study the behavior of   $\epsilon_{k,\text{max}}$ and determine their different scalings as a function of $J$.



\begin{figure}[t]
    \raggedleft
    \includegraphics[width=0.99\linewidth]{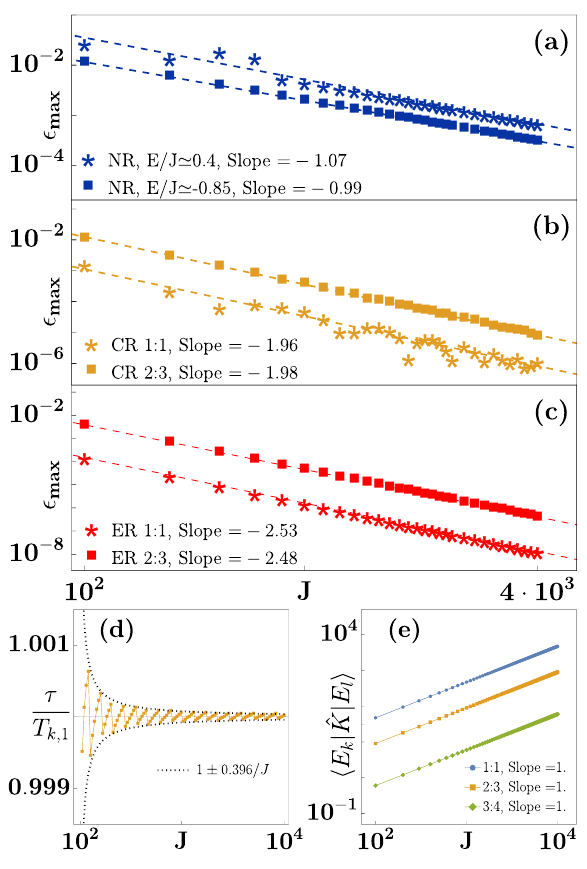}
    \caption{ $\epsilon_{k,\text{max}}$ as a function of $J$ in log-log scale  for (a) non-resonant  states,  (b) states close  to  resonance  and (c) exact resonant states. Numerical fits to the numerical data are indicated in each panel. (d) Time period of the kick over its closest quantum period, $\tau/T_{k,1}$, as a function of $J$ in linear-linear scale. The exact  resonant condition $1{:}1$ is  $\tau/T_{k,1}=1$ (see Sec. IV in \cite{SM} for similar plots for other resonances). (e) Scaling of kick matrix elements for states involved in resonances $1\!:\!1$ ($l=k+1$), $2\!:\!3$ ($l=k+2$) and  $3\!:\!4$ ($l=k+3$).  In panels (a), (b) and (d) the kick period is $\tau=8$. In panel  (c) $\tau$ is adjusted to fulfill exactly the quantum resonant conditions with $\tau\approx 8$. Non-resonant states in (a) are obtained by selecting $\hat{H}_0$ eigenstates closest to the energies  $E/J=0.4$ and $E/J=-0.85$, where, according to Fig.\ref{fig:1}, no resonances are present. Data for (a-c) available at \cite{Raw_data_doi}.
    }
    \label{fig:3}
\end{figure}

\subsection{Scaling of $\epsilon_{k,\text{max}}$.} 
We study the scaling properties of $\epsilon_{k,\text{max}}$ with $J$ under three different conditions: non-resonant NR, close to resonant CR, and exact resonant ER, with a clear distinct scaling behavior of $\epsilon_{k,\text{max}}$. For the latter two, we select the most notable resonances of Fig.\ref{fig:1}, namely 1:1, and 2:3.  As discussed in Sec. IV.D in\cite{SM}, a similar analysis for other resonances would require numerical simulations with much larger $J$ values, which are significantly more computationally expensive. 

The non-resonant condition is obtained by fixing $\tau$ (we use $\tau=8$) and, for each $J$, selecting the state in the $\hat{H}_0$ spectrum that  closely satisfies  $\tau/T_{k,1}=f$, where $f$ is a floating point approximation of an irrational number, in our case we select $f=8/T(\varepsilon)$, where $T(\varepsilon)$ is the classical time period of $h_o$ at the classical energy $\varepsilon\equiv E/J=-0.85$ and $0.4$, where,  according to Fig.\ref{fig:1}, no resonances are present. The condition close to the resonance $m{:}n$ is obtained by fixing $\tau$ and, for each $J$, selecting the state $|E_k\rangle$ in the $\hat{H}_0$ spectrum that comes closest to satisfying the resonant condition 
$\frac{\tau}{n T_{k,m}}=\frac{\tau(E_{k+m}-E_k)}{2\pi n}=1$. Due to the discreteness of the spectrum,  the resonant condition is not fully satisfied, but the ratio $\tau/T_{k,1}$ approaches unity as $J$ increases. This is illustrated in Fig. \ref{fig:3}(d), where it is observed that the ratio $\frac{\tau}{n T_{k,m}}$ oscillates around $1$ with an amplitude that decreases proportionally to $1/J$.  Finally, for the exact  resonant condition, a two-step process is employed. First, we follow a similar approach as in the close to resonance condition to identify $\ket{E_k}$ and $\ket{E_{k+m}}$. Then, for each value of $J$, we fine-tune the kicking period  $\tau$ to exactly satisfy the condition  $\frac{\tau(E_{k+m}-E_k)}{2\pi n}=1$. 

In panels (a)-(c) of Fig. \ref{fig:3}, we show the  scaling of $\epsilon_{k,\text{max}}$ 
for these three distinct conditions: (a) non-resonant, (b) close to resonant, and (c) exact resonant. In all cases, $\epsilon_{k,\text{max}}$ vanishes following a power law, but at different rates. For resonant states,  $\epsilon_{k,\text{max}}$ approaches zero faster than for non-resonant states.
For non-resonant states, we obtain  $$\epsilon_{k,\text{max}}^{\text{NR}}\approx A J^{-1},$$ whereas for close to  resonance and exact resonant states we obtain, respectively,   $$\epsilon_{k,\text{max}}^{\text{CR}}\approx B J^{-2}$$ and   $$\epsilon_{k,\text{max}}^{\text{ER}}\approx C J^{-5/2}.$$ These differentiated   scalings
constitute the main finding of this work. Furthermore, the scalings can be justified by using unitary perturbation theory,  from which, additionally,  we can clearly identify  the roles played by the matrix elements $\hat{K}_{k,k'}\equiv\langle  E_{k}|\hat{K}|E_{k'}\rangle$  and  the quantum resonant condition,  Eq.\eqref{eq:QR}. 

\subsection{ Unitary perturbation theory.} 
Unitary perturbation theory (UPT)  \cite{Peres,Chinni2022} consists of expanding the Floquet quasienergies and (unnormalized) eigenfunctions in powers of the perturbation strength $\epsilon$,
\begin{equation}    \phi_k=\sum_{i=0}\phi_k^{(i)}\epsilon^{i}, \quad \ket{f_k}_u=\sum_{i=0}\epsilon^{i}\ket{f_k^{(i)}}.
\end{equation}

By substituting these expressions in $\hat{F}\ket{f_k}_u=\exp(-i \phi_k)\ket{f_k}_u$, we obtain iteratively the different corrections to the Floquet eigenstates and quasienergies (see Sec. III in \cite{SM}).  At order zero, the quasienergies are given by the eigenphases of $\hat{F}_o=\exp\left(-i\tau \hat{H}_o\right)$, $\phi_{k}^{(0)}=\text{mod}(\tau E_k, 2\pi)$. Similar to standard perturbation theory, the calculation of the perturbative series differs in the degenerate case, which occurs in UPT when  $\phi_{k}^{(0)}=\phi_{k'}^{(0)}$. Due to the modulo $2\pi$, this condition implies $\tau(E_{k}-E_{k'})=2 \pi n$, which is the exact quantum resonant condition in Eq.~\eqref{eq:QR}. Non-degenerate perturbation theory  applies for non-resonant states, as well as states close to   quantum resonance, and degenerate perturbation theory is required to deal with exact quantum resonances.

In the non-degenerate case, after  consistently normalizing the Floquet eigenfunctions, we obtain  (see Sec. IV in \cite{SM}) a perturbative series for the maximum overlap of the Floquet eigenstates (\ref{eq: max epsilon}), 

$    \mathcal{F}_{k,\text{max}} =1-a_2 \epsilon^2-a_3\epsilon^3-a_4\epsilon^4-...$, where 
\begin{equation}
a_2=\langle f_k^{(1)}|f_k^{(1)}\rangle=\frac{1}{4}\sum_{l\neq k}\frac{\abs{
\mel{E_l}{\hat{K}}{E_k}
}^2}{\sin^2\left(\frac{\tau(E_l-E_k)}{2}\right)}.
\label{eq:a2}
\end{equation}
 The matrix elements $\hat{K}_{l,k}\equiv \mel{E_l}{\hat{K}}{E_k}$, as shown in Fig.\ref{fig:3}(e), scale linearly with $J$, implying the numerators in \eqref{eq:a2} scale quadratically with $J$.  Meanwhile, the denominators have relevant  differences depending  on whether the state is non-resonant or close to a resonance. Non-resonant states are free of  small denominators, but  for a state close to a resonance $m{:}n$, the argument of the sine function for $l=k+m$ is very close to a multiple of $\pi$: $$\tau(E_{k+m}-E_k)/2=n \pi+ \delta,$$ where  $\delta$,  as shown in  Fig.\ref{fig:3}(d) ($\frac{\tau}{n T_{k,m}}=1+\frac{\delta}{\pi n}$), approaches zero as $\delta=\delta_{m:n}/J$. Consequently, this resonant sum term yields an additional $J^2$ scaling for $a_2$, coming from $$\frac{1}{\sin^2(\tau (E_{k+m}-E_k)/2)}\approx \frac{J^2}{\delta_{m:n}^2}.$$
 Note that the intensity of a resonance is determined by  the matrix element $\langle E_{k+m}|\hat{K}|E_k\rangle$, and that in order for the resonance to clearly manifest, this matrix  element must be non-zero. 
 From the previous discussion,  we obtain  $$a_2= a_{2,\text{NR}}J^2$$ for non-resonant states, whereas, for close-to-resonant states  $$a_2=a_{2,\text{NR}}J^2+a_{2,\text{CR}} J^4\xrightarrow[]{J\to\infty} a_{2,\text{CR}} J^4.$$
 Similarly, it can be shown (see Sec.~IV.B in \cite{SM}) that for non-resonant states $a_3\approx a_{3,\text{NR}} J^3$, whereas for states close to  resonance the dominant behavior is     $a_3\approx a_{3, \text{CR}} J^6$. If  for large enough $J$ these scalings generalize to higher order perturbative terms, $a_\text{p}=a_{\text{p},\text{NR}}J^{\text{p}}$ and $a_\text{p}=a_{\text{p},\text{CR}}J^{2\text{p}}$, the perturbative series for non-resonant and close to  resonance states are, respectively, 
\begin{eqnarray}
\mathcal{F}_{k,\text{max}}^{\text{NR}}&\approx&1-\sum_{\text{p}=2}^{\text{p}_o}a_{\text{p},\text{NR}} (\epsilon J)^\text{p}\\
\mathcal{F}_{k,\text{max}}^{\text{CR}}&\approx&1-\sum_{\text{p}=2}^{\text{p}_o}a_{\text{p},\text{CR}} (\epsilon J^2)^\text{p}.
\end{eqnarray}
To determine $\epsilon_{k,\text{max}}$, we solve the equation $\mathcal{F}_{k,\max}=1/2$, Eq.~\eqref{eq: max epsilon}, for $\epsilon$. By truncating the perturbative series until a certain power  of $\epsilon$, $\text{p}_o$, we are left with a polynomial equation in terms of  variable $z=\epsilon J$ for non-resonant states and in terms of  variable $z=\epsilon J^2$ for states close to resonance. Let  $z_{\text{NR}}$ and $z_{\text{CR}}$ 
denote the respective physically meaningful solutions of these equations. Then, for non-resonant states, we obtain $\epsilon_{k,\text{max}}=z_{\text{NR}}/J$, whereas  for states  close to  resonance $\epsilon_{k,\text{max}}=z_\text{CR}/J^2$, which are the scalings     obtained  numerically, shown in Fig.\ref{fig:3}(a)-(b).    

For exact resonances we employ degenerate perturbation theory (see Sec. III.B in \cite{SM}). In this case, the degenerate subspace of $\hat{F}_o$ is  spanned by the states $|E_{k}\rangle$ and $|E_{k+m}\rangle$. The zeroth order corrections to the Floquet eigenfunctions and first order corrections to the quasi-energies, $\phi_{a}^{(1)}$, are obtained from diagonalizing the kick operator  in this  subspace 
\begin{equation}
\left(\begin{array}{cc}
\hat{K}_{k,k}    & \hat{K}_{k,k+m} \\
\hat{K}_{k+m,k}    &\hat{K}_{k+m,k+m} 
\end{array}\right)
\left(\begin{array}{c}
     c_{a,1}\\
     c_{a,2}      
\end{array}\right)= \phi^{(1)}_{a}\left(\begin{array}{c}
     c_{a,1}\\
     c_{a,2}      
\end{array}\right),
\label{eq:deg}
\end{equation}
  with $a=k,k+m$ and,   assuming $|c_{k,1}|>|c_{k,2}|$,
\begin{eqnarray}
    |f_{k}^{(0)}\rangle&=& c_{k,1} |E_k\rangle+c_{k,2} |E_{k+m}\rangle,\\
    |f_{k+m}^{(0)}\rangle&=& c_{k+m,1} |E_k\rangle+c_{k+m,2} |E_{k+m}\rangle.
\end{eqnarray}
By analyzing the scaling of the matrix elements  in \eqref{eq:deg}, we obtain (see Sec.~V in \cite{SM}) $$c_{k,1}\approx\frac{1}{\sqrt{2}}\left(1+ \frac{c}{J}\right).$$ This implies that in the case of exact resonance, 
even for infinitesimal $\epsilon$, the eigenstates of the Floquet operator are a linear combination of two eigenstates of $\hat{H}_o$ with almost the same  contribution from each. The perturbative expression for the maximum overlap until quadratic terms  is now 
$$\mathcal{F}_{k,\text{max}}^{\text{ER}}=(1-a_2^{\text{ER}} \epsilon^2)|c_{k,1}|^2,$$
where $a_2^{\text{ER}}$ is  given by the same expression as in  Eq.\eqref{eq:a2} with $|E_k\rangle\rightarrow|f_k^{(0)}\rangle$ and the sum restriction  now excluding  the two states involved in the exact resonance~ ($\sum_{l\not=k,k+m}$), thus avoiding  an ill-defined expression. From the scalings of the matrix elements and small denominators, we  obtain $a_2^{\text{ER}}=a_{2,\text{ER}}
J^4$, which implies $$\mathcal{F}_{k,\text{max}}^{\text{ER}}=\frac{1}{2}+\frac{c}{J}-\frac{a_{2,\text{ER}}}{2} J^4 \epsilon^2.$$ 
Then, from $\mathcal{F}_{k,\text{max}}^{\text{ER}}=\frac{1}{2}$, we deduce that     $$\epsilon_{k,\text{max}}=\sqrt{\frac{2c}{a_{2,\text{ER}}}} \frac{1}{J^{5/2}},
$$ which coincides very well with the scaling obtained numerically,  shown in Fig.\ref{fig:3}(c). 

The analysis in this subsection demonstrates that, for sufficiently large $J$, the distinct sensitivity to perturbations in both non-resonant and resonant states can be  described using unitary perturbation theory.



\section{Conclusions} 
\label{Sec:Discussion}
By studying the eigenstates of the Floquet operator in terms of the eigenstates of the unperturbed Hamiltonian $\hat{H}_0$ in a kicked spin model for large but finite $J$, we have elucidated a mechanism which renders resonance states fragile  and non-resonance states robust against perturbation. The results are expected to hold under general periodic driving, as the resonance conditions depend only on the time periodicity and not on the specific form of the driving. We  combine numerically exact results with analytical expressions obtained from unitary perturbation theory to derive preliminary steps toward a quantum counterpart of the classical  KAM theorem. A  potential application of our work is to provide a comprehensive framework that elucidates the effects of classical KAM resonances on quantum phenomena, such as measurement-induced transmon ionization. This has been explored in a recent study \cite{Blais2024PRX} where classical KAM resonances have been leveraged to explain the observed ionization in circuit quantum electrodynamics. Additionally, our framework could enhance the understanding of time crystals in the kicked LMG model \cite{KlmgCrystalPRB2017,Poggi2022PRR,YaoCrystalReview2023} and other types of resonances in kicked systems \cite{VarikutiPRE2024,Wang2013PRE,Ramareddy2010,Casati2000}. It 
also  suggests an alternative approach to quantum state engineering \cite{bimbard2010quantum}  across various experimental platforms \cite{bordia2017periodically,ringot2000,venkatraman2024driven}. The effect of resonances in the ground-state or, more generally, in states of $\hat{H}_0$ associated with classical fixed-points is an interesting extension of the study presented here.

\begin{acknowledgments}
MK, SH, and AST would like to thank Lauren Hayward and Ayana Sarkar for helpful discussions. MK acknowledges the support from the Applied Quantum Computing Challenge
Program at the National Research Council of Canada. This research was supported in part by Perimeter Institute for Theoretical Physics. Research at Perimeter Institute is supported in part by the Government of Canada through the Department of Innovation, Science and Economic Development and by the Province of Ontario through the Ministry of Colleges and Universities. JAS-L is grateful to the people of Mexico, who through Secretar\'ia de Ciencias, Humanidades, Tecnolog\'ia e Innovaci\'on (SECIHTI) funded his graduate education,  CVU number:1181841. We acknowledge the support of the Computational Center ICN-UNAM, in particular to Enrique Palacios, Luciano Diaz, and Eduardo Murrieta. DJN acknowledges financial support from grant CZ.02.01.01/00/22\_008/0004649
(QUEENTEC) provided by  MEYS of the Czech Republic. 
\end{acknowledgments}
\bibliographystyle{unsrt}
\input{LetterWO-GS_v3.bbl}

\end{document}


\title{Supplemental material: Resonant fragility and nonresonant robustness of Floquet eigenstates in kicked spin systems
}
\author{Jes\'us A. Segura-Landa \textsuperscript{1,2}}
\author{Meenu Kumari\textsuperscript{3,4}}
\author{Daniel J. Nader\textsuperscript{5}}
\author{Sercan H\"usn\"ugil\textsuperscript{4,6}}
\author{Ali SaraerToosi\textsuperscript{4,7}}
\author{Sergio Lerma-Hern\'andez\textsuperscript{1}}
 \affiliation{\textsuperscript{1}Facultad  de F\'isica, Universidad Veracruzana,  Campus Arco Sur, Paseo 112, C.P. 91097  Xalapa, Mexico.}
 \affiliation{\textsuperscript{2}Instituto de Ciencias Nucleares, Universidad Nacional Aut\'onoma de M\'exico, Apdo. Postal 70-543, C.P. 04510  Cd. Mx., Mexico}
\affiliation{\textsuperscript{3}Digital Technologies, National Research Council Canada}
\affiliation{\textsuperscript{4}Perimeter Institute for Theoretical Physics, Waterloo ON N2L 2Y5, Canada}
\affiliation{\textsuperscript{5}Department of Optics, Faculty of Science, Palacky University, Olomouc, 77900, Czech Republic}
\affiliation{\textsuperscript{6}Department of Physics and Astronomy, University of Waterloo, Waterloo, Ontario, N2L 3G1, Canada}
\affiliation{\textsuperscript{7}Department of Computer Science, University of Toronto, 40 St. George St., Toronto, ON, M5S 2E4, Canada}
\maketitle
\tableofcontents
\section{Nonlinear resonances and classical time period }

In classical physics, all time-independent Hamiltonians with one degree of freedom are integrable, thus expressible as $h_o = h_o(I)$ in the action-angle variable form. Hence, we study a one-degree-of-freedom time-periodic Hamiltonian to explore integrability breaking, given by:

\begin{equation}
    h(I,\theta,t) = h_o(I) + \epsilon k(I,\theta,t), \quad \epsilon \ll 1,
\end{equation}
where $h_o(I)$ represents the integrable Hamiltonian with action-angle variables \((I, \theta)\) and $k(I,\theta,\tau)$ is a time-periodic perturbing Hamiltonian with period $\tau$, i.e., $   k(I,\theta,t+\tau) = k(I,\theta,t)$. For the integrable Hamiltonian considered in the main text, the classical hamiltonian is obtained from Bloch coherent states, $|\alpha\rangle$, as
\begin{equation}\label{SM h_o}
   h_o(\phi,z_c)\equiv \lim_{J\to\infty}\frac{\mel{\alpha}{\hat{H}_0}{\alpha}}{J}= z_c+\frac{(1-z_c^2)}{2}[\gamma_x\cos^2\phi+\gamma_y\sin^2\phi],
\end{equation}
where $\langle \alpha|\hat{J}_z|\alpha\rangle/J= z_c$, $ \langle\alpha|  \hat{J}_x|\alpha\rangle/J=\sqrt{1-z_c^2}\cos\phi$ and $ \langle\alpha|  \hat{J}_y|\alpha\rangle/J=\sqrt{1-z_c^2}\sin\phi$. The action of the torus with the rescaled energy $\varepsilon\equiv E/J$ is defined as $I=\frac{1}{2\pi}\oint z_c(\varepsilon,\phi)d\phi$. Here we have used the canonical coordinates $(\phi,z_c)$, which are related to the canonical coordinates used in main text, $(Q,P)$, through  $Q=\sqrt{2(1+z_c)}\cos\phi$ and $P=-\sqrt{2(1+z_c)}\sin\phi$. A resonance between the time degree of freedom of $k(I,\theta,t)$ and the degree of freedom of $h_o$ occurs whenever the driving frequency of the perturbing Hamiltonian $\omega = 2\pi/\tau$ matches a rational multiple of the natural frequency of $h_o$, given  by $\Omega(I) = dh_o/dI$: 

\begin{equation}\label{eq: supmat resonant condition}
    m\Omega_{m:n} = n\omega, \quad \Omega_{m:n} = \left.\left(\frac{dh_o}{dI}\right) \right|_{I=I_{m:n}}, \quad m,n \in \mathbb{Z},
\end{equation}
where $I_{m:n}$ is the action of the torus whose frequency is associated with the $m{:}n$ resonance. For a tori with energy $\varepsilon$ in the classical phase space of $h_o$, the time period $T(\varepsilon)$ of the tori is given by (note that all tori of $h_o(I)$ are closed orbits): 
\begin{equation}
    T(\varepsilon) = \frac{2\pi}{\Omega(\varepsilon)} = 2\pi \frac{dI(\varepsilon)}{d\varepsilon}.
\end{equation}
In terms of time periods, the resonant condition \eqref{eq: supmat resonant condition} is written as
\begin{equation}
    m\,\tau=n\,T(I_{m:n}),
\end{equation}
and it generates an $m-$cycle in the stroboscopic dynamics in the phase space. 
The classical time period for an orbit of energy $\varepsilon$ of the Hamiltonian \eqref{SM h_o} is obtained from the following integral 

\begin{equation}
    T(\varepsilon)=\int\limits_{0}^{2\pi} \frac{d\phi}{\sqrt{1-2\varepsilon A(\phi)+A(\phi)^2}},
    \label{eq:ClPeriod}
\end{equation}
for $|\gamma_x|,|\gamma_y| \leq 1$, where $A(\phi)=\gamma_x\cos^2(\phi)+\gamma_y\sin^2(\phi)$. For the parameters used in the main text, $\gamma_y=0$ and $\gamma_x=-0.95$, it reduces to $A(\phi)=-0.95\cos^2(\phi)$. The  general expression for the  classical  LMG Hamiltonian can be found in Ref.  \cite{NaderAvoided} of main text. For the time-dependence of $k(I,\theta,\tau)$, we choose time-periodic delta function, $\delta(t-n\tau)$, $n \in \mathbb{Z}$ which makes the analytical calculations for the model in quantum physics possible using unitary perturbation theory. As the time-periodic delta function imparts a periodic kick to the system, we refer to the driving time period and driving frequency as kicking period and kicking frequency, respectively.

\section{Participation ratio for exact quantum resonances}

\begin{figure}
    \centering    
        \includegraphics[width=1.0\linewidth]{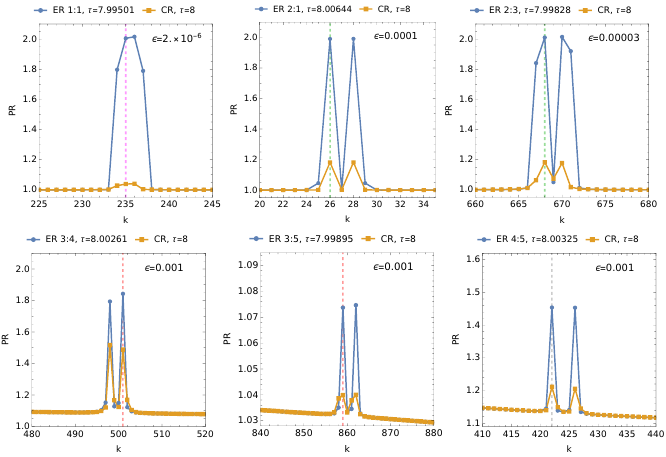}
    \caption{Participation ratio (PR) of the Floquet eigenstates, $\ket{f_k}$, in the energy eigenbasis, $\ket{E_n}$ of the LMG model. The top row shows PR for resonances $1\!:\!1$, $2\!:\!1$ and $2\!:\!3$ from left to right. The bottom row shows PR for resonances $3\!:\!4$, $3\!:\!5$ and $4\!:\!5$. In all cases, $J=500$, blue data points indicate the PR for exact quantum resonances (with the kicking period adjusted for each case), and orange data points represent approximate quantum resonance with $\tau=8$. The perturbation strength, $\epsilon$, of the kick is specified in each plot.}
    \label{fig: SM ExactResonances}
\end{figure}

The participation ratio (PR) of normalized Floquet eigenstates, $\ket{f_k}$, in the unperturbed Hamiltonian eigenbasis, $\ket{E_n}$, is defined as 
\begin{equation}
\text{PR}\equiv\left(\sum_{n=1}^{N}\abs{\braket{E_n}{f_k}}^{4}\right)^{-1},
\end{equation}
where $N$ is the Hilbert space dimension (for pseudo-spin systems, $N=2J+1$). Also referred as the number of principal components (NPC) in literature \cite{YurovskyPRL2023}, the PR measures how extensively the state $\ket{f_k}$ is spread in the eigenbasis $\ket{E_n}$. Its minimum value is equal to $1$ when $\ket{f_k}$ coincides with a single state of the basis $\ket{E_n}$, and maximum value is $N$ when   $\ket{f_k}$ is evenly distributed across all the elements of the  eigenbasis  $\ket{E_n}$. In contrast, the inverse of the Participation Ratio (IPR), defined as $\text{IPR}=1/\text{PR}$, is maximal (equal to 1) when the state is maximally localized, and minimal (equal to $1/N$) when the state is fully delocalized. 

The PR is highly sensitive to the proximity of resonances, showing sharp peaks for states close to resonances (as shown in Fig.~\ref{fig:1}(b) of main text where $\ln(\text{PR})$ is plotted for $\ket{f_k}$), as well as for states involved in exact quantum resonances. States involved in exact quantum resonances show  a pronounced increase  of the PR, as shown in Fig.~\ref{fig: SM ExactResonances}, where we plot the PR for close-to-resonance (CR) states (orange) and exact resonance (ER) states (blue) for resonances $1\!:\!1$, $2\!:\!1$, $2\!:\!3$, $3{:}4$, $3{:}5$ and $4{:}5$. While the kicking period was fixed for all the CR states to $\tau = 8$, this period was fine-tuned for the case of ER states as mentioned in the figures. In all cases, the exact resonant condition considerably increases the PR of resonant states, indicating greater delocalization of the involved states.


\section{Unitary perturbation theory}
In this section, we review  the unitary perturbation theory \cite{Peres}  and present  expressions for the lower-order corrections. Similar to standard perturbation theory, we distinguish  between   non-degenerate and degenerate cases.   

Unitary perturbation theory enables us to derive approximate expressions for Floquet eigenstates and quasienergies provided that the Floquet operator $\hat{F}$ can be expressed as a product of two unitaries: $\hat{F}=\hat{F}_0 \hat{F}_K=\exp\left(-i\tau \hat{H}_0\right)\exp\left(-i\epsilon \hat{K}\right)$. This form arises when the time-dependence in the perturbing Hamiltonian is a periodic delta function. It also requires the knowledge of the eigenfunctions and eigenvalues  of $\hat{H}_0$, $\hat{H}_0|E_k\rangle=E_k|E_k\rangle$ with $\langle E_{k'}|E_k\rangle=\delta_{k' k}$. Consequently, $\hat{F}_0 |E_k\rangle=\exp(-i\tau E_k)|E_k\rangle=\exp\left[-i\!\!\! \mod\!(\tau E_k,2\pi)\right]|E_k\rangle$. To obtain the approximate expressions, the  quasienergies and  Floquet eigenstates are expanded in powers of $\epsilon$:
\begin{equation}
    \begin{split}
        \phi_{k}&=\sum_{i=0} \phi_{k}^{(i)} \epsilon^i
        =\phi_k^{(0)}+\epsilon \phi_{k}^{(1)}+\epsilon^2 \phi_k^{(2)}+...
    \end{split}
\end{equation}
\begin{equation}\label{eq: fk}
    \begin{split}
        |f_{k}\rangle_u&=\sum_{i=0} \epsilon^i |f_{k}^{(i)}\rangle
        = |f_k^{(0)}\rangle+\epsilon^{1}|f_k^{(1)}\rangle+\epsilon^{2}|f_k^{(2)}\rangle+...
    \end{split}
\end{equation}
We consider that the zero-order corrections form an orthonormal basis,  $\langle f_{k'}^{(0)}|f_{k}^{0}\rangle=\delta_{k'k}$, and without loss of generality it can be assumed that the higher order corrections are orthogonal to them, $ \langle f_{k}^{(0)}|f_k^{(i)}\rangle=0 $  for $i\geq 1$. These assumptions imply that the eigenfunctions $|f_{k}\rangle_u$ are unnormalized. Normalized eigenfunctions can be obtained by expanding consistently in powers of $\epsilon$ the expression $|f_k\rangle= Z |f_k\rangle_u$, where the normalization constant is $Z=\frac{1}{\sqrt{ _u\langle f_k|f_k\rangle_u}} $.

After expanding  $\hat{F}$ in powers of $\epsilon$, 
we substitute 
the perturbative series into the equation $\hat{F}|f_k\rangle_u=\exp\left(-i\phi_k\right)|f_k\rangle_u$, 
\begin{align}
 &e^{-i \hat{H}_0\tau}\left( |f_k^{(0)}\rangle+ \epsilon \left[ |f_k^{(1)}\rangle-i\hat{K} |f_k^{(0)}\rangle\right]+ \epsilon^2 \left[|f_k^{(2)}\rangle- i\hat{K}|f_k^{(1)}\rangle-\frac{1}{2}\hat{K}^2|f_k^{(0)}\rangle  \right]+...\right)= \label{eq:perser}\\
&e^{-i\phi_k^{(0)}}\left(|f_k^{(0)}\rangle+\epsilon\left[ |f_k^{(1)}\rangle-i\phi_{k}^{(1)}|f_k^{(0)}\rangle\right]+\epsilon^2\left[|f_k^{(2)}\rangle-i\phi_k^{(1)}|f_k^{(1)}\rangle- \left\{i\phi_k^{(2)}+\frac{1}{2}\left(\phi_{k}^{(1)}\right)^2 \right\}|f_k^{(0)}\rangle \right]+... \right). \nonumber
\end{align}
In this way, we obtain a set of equations by   equating the terms with the same power of $\epsilon$. Then, we use 
the resulting equations 
to obtain iteratively the successive corrections to the quasienergies and Floquet eigenstates, as we explain in the following subsections for the non-degenerate and degenerate cases, respectively.

\subsection{Non-degenerate case}

For an eigenstate of $\hat{F}_0=e^{-i\hat{H}_0\tau}$, whose quasienergy is non-degenerate, $\phi_k^{(0)}=\mod(\tau E_k,2\pi) \neq \mod(\tau E_{k'},2\pi) $ for $k'\not=k$, which is the case for non-resonant states and also for close-to-resonance quantum states, 
the zero-order equation,
$$
e^{-i \hat{H}_0\tau} |f_k^{(0)}\rangle=e^{-i\phi_k^{(0)}}|f_k^{(0)}\rangle,
$$
implies that  $|f_k^{(0)}\rangle=|E_k\rangle$. From this, the first order equation becomes
$$
e^{-i \hat{H}_0\tau} \left[ |f_k^{(1)}\rangle-i\hat{K} |E_k\rangle\right]= e^{-iE_k\tau}\left[ |f_k^{(1)}\rangle-i\phi_{k}^{(1)}|E_k\rangle\right].
$$
We multiply this equation by $\langle E_k| $, and use $\langle E_k|f_k^{(1)}\rangle=0$ and $\langle E_k|E_k\rangle=1$ to obtain $ 
\phi_{k}^{(1)}=\langle E_k|\hat{K}|E_k\rangle$. The  first order correction to the Floquet eigenstates are obtained from  multiplying  the same equation but now by $\langle E_{k'}| $  with $k'\not= k$,
\begin{equation}
    \left(e^{-i E_{k'}\tau}-e^{-i E_k\tau}\right)\langle E_{k'}|f_k^{(1)}\rangle=i e^{-i E_{k'}\tau}\langle E_{k'}|\hat{K}|E_k\rangle. \ 
\end{equation}
From this, we obtain the components of the first order correction to the Floquet eigenstates  in the $\hat{H}_0$ eigenbasis, $\langle E_{k'}|f_k^{(1)}\rangle$. By repeating this procedure for higher order equations,  we obtain the  corrections to the quasienergies 
\begin{eqnarray}
\phi_{k}^{(0)}&=& E_k \tau\\
 \phi_{k}^{(1)}&=& \langle E_k| \hat{K}| E_k \rangle\\
  \phi_k^{(2)}&=&\langle E_k| \hat{K}| f_k^{(1)} \rangle-\frac{i}{2}\langle E_k| \hat{K}^2| E_k \rangle+\frac{i}{2}\left(\phi_{k}^{(1)}\right)^2
  =\frac{1}{2} \sum_{k'\not=k} \cot\left[\frac{\tau(E_k-E_{k'})}{2}\right] \left|\langle E_k| \hat{K}| E_{k'} \rangle\right|^2\\    \phi_{k}^{(3)}&=&i\phi_k^{(1)}\phi_k^{(2)}+\frac{1}{6}\left(\phi_k^{(1)}\right)^3+ \langle E_{k}| \hat{K}| f_k^{(2)} \rangle-\frac{i}{2} \langle E_{k}| \hat{K}^2| f_k^{(1)} \rangle-\frac{1}{6} \langle E_{k}| \hat{K}^3| E_k \rangle.   
\end{eqnarray}
Subsequent to the quasienergy corrections, the components of the corrections to the unnormalized Floquet eigenfunctions, in the $\hat{H}_0$ eigenbasis, are given by ($k'\not = k$ for corrections of order $i \geq 1$)
\begin{eqnarray}
    \langle E_{k'}|f_k^{(0)}\rangle&=& \delta_{k' k} \label{eq:f0}\\  
    \langle E_{k'}|f_k^{(1)}\rangle&=& -i\frac{\langle E_{k'}| \hat{K}| E_k \rangle} {e^{i\tau(E_{k'}-E_k)}-1} \label{eq:f1} \\
   \left(e^{i\tau(E_{k'}-E_k)}-1\right)  \langle E_{k'}|  f_k^{(2)} \rangle&=&\!\!\!-i\langle E_{k'}| \hat{K}| f_k^{(1)} \rangle  -\frac{1}{2} \langle E_{k'}| \hat{K}^2| E_k \rangle+ 
   i e^{i\tau(E_{k'}-E_k)}\phi_k^{(1)} \langle E_{k'}|  f_k^{(1)} \rangle \label{eq:f2}\\
     \left(e^{i\tau(E_{k'}-E_k)}-1\right)  \langle E_k'|  f_k^{(3)} \rangle&=&-i \langle E_{k'}| \hat{K}| f_k^{(2)} \rangle-\frac{1}{2}\langle E_{k'}| \hat{K}^2| f_k^{(1)} \rangle+\frac{i}{6}\langle E_{k'}| \hat{K}^3| E_k \rangle+\label{eq:f3}\\
   & &e^{i\tau(E_{k'}-E_k)}\left( i \phi_k^{(1)} \langle E_{k'}|  f_k^{(2)} \rangle + \left[ i \phi_k^{(2)}+\frac{1}{2} \left(\phi_k^{(1)}\right)^2\right] \langle E_{k'}|  f_k^{(1)} \rangle\right). 
  \nonumber
\end{eqnarray}
Moreover, if the state $|E_k\rangle$ satisfies an exact quantum resonant condition $\tau(E_{k'}-E_k)= 2 \pi n$, where $n\in \mathbb{Z}$ and $k'\not= k$, the previous  expressions contain ill-defined terms. As is done in standard perturbation theory, a modified version of the perturbative expressions must be used in this scenario, which is discussed in the following subsection. 

\subsection{Degenerate case}
In unitary perurbation theory,  we have a degeneracy when two (or more) quasienergies of $\hat{F}_0$ are equal,  $\phi_{k}^{(0)}=\text{mod}(\tau E_k,2\pi)=\phi_{k'}^{(0)}=\text{mod}(\tau E_{k'},2\pi)$, for $k'\not= k$. Note that this can happen even when the spectrum of $\hat{H}_0$ is non-degenerate. 

Suppose a two-fold degeneracy between states $|E_k\rangle$ and $|E_{k+m}\rangle$, and $\phi_k^{(0)}=\mod(E_k \tau,2\pi)=\mod(E_{k+m} \tau,2\pi)=\phi_{k+m}^{(0)}$. Due to the modulo $2\pi$, we have a degeneracy for states $|E_k\rangle$ and $|E_{k+m}\rangle$ if $$
\tau(E_{k+m}-E_{k})=2\pi n, \ \ \ \ \ \ \ \ \ n \in \mathbb{Z},
$$ 
which is the resonant condition discussed in the main text,  Eq.~\eqref{eq:QR}.

Then, from the zero-order equation of Eq.~\eqref{eq:perser},
$$
e^{-i \hat{H}_0\tau} |f_a^{(0)}\rangle=e^{-i\phi_a^{(0)}}|f_a^{(0)}\rangle,
$$
with $a=k,k+m$, we  have two orthonormal states
\begin{eqnarray}
    |f_{k}^{(0)}\rangle&=& c_{k,1} |E_k\rangle+c_{k,2} |E_{k+m}\rangle \label{eq:fa}\\
    |f_{k+m}^{(0)}\rangle&=& c_{k+m,1} |E_k\rangle+c_{k+m,2} |E_{k+m}\rangle,\label{eq:fap}
\end{eqnarray}
where  the coefficients $c_{a,i}$ have to be determined, but in  order to label the states $|f_{a}^{(0)}\rangle$, we have assumed that $|c_{k,1}|\geq|c_{k,2}|$, which implies  $|c_{k+m,1}|\leq|c_{k+m,2}|$ because of orthogonality of $|f_{k}^{(0)}\rangle$ and $|f_{k+m}^{(0)}\rangle$.
From the first order equation of Eq.\eqref{eq:perser}, 
\begin{equation}
e^{-i \hat{H}_0\tau} \left[ |f_a^{(1)}\rangle-i\hat{K} |f_a^{(0)}\rangle\right]= e^{-i\phi_a\tau}\left[ |f_a^{(1)}\rangle-i\phi_{a}^{(1)}|f_a^{(0)}\rangle\right],\label{eq:firstdeg}
\end{equation}
we obtain, after multiplying it by $\langle f_a^{(0)}|$, 
\begin{equation}
    \phi_{a}^{(1)}=\langle f_a^{(0)}|\hat{K}|f_{a}^{(0)}\rangle.\label{eq:diagdeg}
\end{equation}

On the other hand, if  we now  multiply Eq.\eqref{eq:firstdeg} by $\langle \phi_{a'}^{(0)}|$, with $a'=k,k+m$, $a'\not= a$, we obtain
\begin{equation}
  \langle f_{a'}^{(0)}|\hat{K}|f_a^{(0)}\rangle=ie^{i\phi_{a'}}\left( e^{-i\phi_{a}^{(0)}}-e^{-i \phi_{a'}^{(0)}}\right)\langle f_{a'}^{(0)}|f_{a}^{(1)}\rangle=0,
  \label{eq:nodiagdeg}
\end{equation}
where in the last equality we have used the degeneracy $\phi_{a}^{(0)}=\phi_{a'}^{(0)}$. 
Eqs. \eqref{eq:diagdeg} and \eqref{eq:nodiagdeg} imply that the zeroth order  eigenfunctions of the Floquet operator  and first order corrections to the quasienergies  are obtained from  diagonalizing the perturbation $\hat{K}$ in the degenerate subspace, i.e., from   the  normalized  eigenvectors of the hermitian matrix
\begin{equation}
\left(\begin{array}{cc}
\langle E_{k}|\hat{K}|E_{k}\rangle     & \langle E_{k}|\hat{K}|E_{k+m}\rangle \\
\langle E_{k+m}|\hat{K}|E_{k}\rangle     &\langle E_{k+m}|\hat{K}|E_{k+m}\rangle 
\end{array}\right)\left(\begin{array}{c}
     c_{a,1}\\
     c_{a,2}      
\end{array}\right)= \phi_{a}^{(1)}\left(\begin{array}{c}
     c_{a,1}\\
     c_{a,2}      
\end{array}\right),
\label{eq:degem}
\end{equation}
where $a=k,k+m$ and the coefficients $c_{a,i}$ determine $|f_{a}^{(0)}\rangle$ according to Eqs.\eqref{eq:fa} and \eqref{eq:fap}. 

The components of the  first-order correction to the unnormalized eigenfunctions,  $|f_a\rangle_u$, in the $\hat{H}_0$ eigenbasis for $k'\not=k,k+m$  are obtained from multiplying  Eq.\eqref{eq:firstdeg}  by the bra $\langle E_{k'}|$ 
\begin{equation}
    \langle E_{k'}|f_{a}^{(1)}\rangle=-i\frac{\langle E_{k'} |\hat{K}|f_{a}^{(0)}\rangle}{e^{i\tau (E_{k'}-E_{a})}-1} \ \ \ \  \ \ \text{for $k'\not=k, k+m$}.
    \label{eq:firstW}
\end{equation}
It remains to calculate the component of the first order correction in the degenerate subspace, $\langle f_{a'}^{(0)}|f_{a}^{(1)}\rangle$
for $a,a'=k$ or $k+m$ ($a'\not= a$). This component is obtained from the $\epsilon^2$ terms in Eq.\eqref{eq:perser},  
\begin{align}
&e^{-i \hat{H}_0\tau}\left[|f_a^{(2)}\rangle- i\hat{K}|f_a^{(1)}\rangle-\frac{1}{2}\hat{K}^2|f_a^{(0)}\rangle  \right]=\\
&e^{-i\phi_a^{(0)}}\left[|f_a^{(2)}\rangle-i\phi_a^{(1)}|f_a^{(1)}\rangle- \left\{i\phi_a^{(2)}+\frac{1}{2}\left(\phi_{a}^{(1)}\right)^2 \right\}|f_a^{(0)}\rangle \right].\nonumber
\end{align}
After multiplying it by $\langle f_{a'}^{(0)}|$ with $a'\not=a$, using the degeneracy condition, $\phi_{a'}^{(0)}=\phi_a^{(0)}$, and the orthogonality  $\langle f_{a'}^{(0)}|f_{a}^{(0)}\rangle=0$, we obtain
$$
 \phi_{a}^{(1)}\langle f_{a'}^{(0)}|f_{a}^{(1)}\rangle=  \langle f_{a'}^{(0)}|\hat{K}|f_{a}^{(1)}\rangle-\frac{i}{2}\langle f_{a'}^{(0)}|\hat{K}^2|f_{a}^{(0)}\rangle.
$$
Then, we use the resolution of the identity $\mathbf{1}=\displaystyle\sum_{k'\not=k,k+m}|E_{k'}\rangle \langle E_{k'}| + | f_a^{(0)}\rangle\langle f_{a}^{(0)}|+| f_{a'}^{(0)}\rangle\langle f_{a'}^{(0)}|$ in the first term of the RHS of the previous equation to get
$$
 \phi_{a}^{(1)}\langle f_{a'}^{(0)}|f_{a}^{(1)}\rangle=  \langle f_{a'}^{(0)}|\hat{K}|f_{a'}^{(0)}\rangle\langle f_{a'}^{(0)}|f_a^{(1)}\rangle+\sum_{k'\not=k,k+m}\langle f_{a'}^{(0)}|\hat{K}|E_{k'}\rangle\langle E_{k'}|f_a^{(1)}\rangle- \frac{i}{2}\langle f_{a'}^{(0)}|\hat{K}^2|f_{a}^{(0)}\rangle,
$$
where we have used $\langle f_{a'}^{(0)}|\hat{K}|f_{a}^{(0)}\rangle=0$. Finally, substituting Eqs.\eqref{eq:diagdeg} and \eqref{eq:firstW} in the previous equation, we obtain 
\begin{equation}
    \langle f_{a'}^{(0)}|f_{a}^{(1)}\rangle=-\frac{i}{\phi_{a}^{(1)}-\phi_{a'}^{(1)}}\left(\sum_{k'\not= k,k+m}\!\!\!\!\! \frac{\langle f_{a'}^{(0)}|\hat{K}|E_{k'} \rangle\langle E_{k'}|\hat{K}| f_{a}^{(0)}\rangle}{e^{i\tau(E_{k'}-E_{a})}-1} +\frac{1}{2}\langle f_{a'}^{(0)}|\hat{K}^2|f_{a}^{(0)}\rangle\right).
    \label{eq:fofu}
\end{equation}
 Higher order corrections can be obtained similarly, but  for our purposes these corrections are sufficient. 
 
Similar to the corrections for the non-degenerate case, the expressions for this degenerate case may have small denominators, as discusses in detail in Sec.\ref{sec:ApPFER}, but in contrast to the non-degenerate case, the previous expressions have no ill-defined terms when the  resonant condition is exactly fulfilled. 


\section{Perturbative series for $\mathcal{F}_\text{max}$: non-resonant and close-to-resonance cases}

In this section, we derive the  perturbative series of $\mathcal{F}_{k,\text{max}}$ for a non-degenerate case and determine the scaling of its terms with respect to the system size, $J$. This analysis allows us to establish  the scaling of  $\epsilon_{\text{max}}$ for non-resonant and close-to-resonance cases, as  discussed in the main text.

For small $\epsilon$ in unitary perturbation theory, the maximum overlap of the Floquet eigenstates with respect to the $\hat{H}_0$ eigenbasis, $\mathcal{F}_{k,\text{max}}\equiv\max\limits_{n}\abs{\braket{f_k}{E_{n}}}^2$ is simply given by $\mathcal{F}_{k,\text{max}}=\abs{\langle f_k|E_{k}\rangle }^2$, where  $|f_k\rangle$ is the normalized perturbative approximation to the Floquet eigenstate and $|E_k\rangle$ is the eigenstate of $\hat{H}_o$ with the same index $k$ in the perturbative expansion.

For non-degenerate perturbation theory, the normalized Floquet eigenstate is given by $$|f_k\rangle=Z|f_{k}\rangle_u=Z\left(|E_{k}\rangle+\epsilon |f_k^{(1)}\rangle+ ..\right),$$ where $Z$ is the normalization constant, $Z=\frac{1}{\sqrt{ _u\langle f_k|f_k\rangle_u}}$. Since $\langle E_{k}|f_{k}^{(i)}\rangle=0$ for $i\geq 1$, the maximum overlap is given entirely by the squared normalization constant, $$\mathcal{F}_{k,\text{max}}=|\langle f_k|E_k\rangle|^2=|Z|^2=\frac{1}{ _u\langle f_k|f_k\rangle_u}=\frac{1}{1+ \langle f_k^{(1)}|f_k^{(1)}\rangle \epsilon^2+ 2 \text{Re}\left[\langle f_k^{(1)}|f_k^{(2)}\rangle \right]  \epsilon^3 +...}. $$

By expanding the previous equation in powers of $\epsilon$,  
we obtain until fourth order 
\begin{align}    \mathcal{F}_{k,\text{max}}&= 1-a_2 \epsilon^2-a_3 \epsilon^3-a_4 \epsilon^4-...
  \label{eq:fmax4}   \\ &=1- \langle f_k^{(1)}|f_k^{(1)}\rangle\epsilon^2-2 \text{Re}\left[\langle f_k^{(1)}|f_k^{(2)}\rangle \right]  \epsilon^3 -
     \left(2 \text{Re}\left[\langle f_k^{(1)}|f_k^{(3)}\rangle\right]+  \langle f_k^{(2)}|f_k^{(2)}\rangle- \langle f_k^{(1)}|f_k^{(1)}\rangle^2\right)\epsilon^4-...\nonumber
\end{align}
The fact that $\mathcal{F}_{k,\text{max}}$ does not contain terms of $\order{\epsilon^1}$ can be understood from the fact that $\langle E_k|f_k^{(1)}\rangle = 0$.

\subsection{Second order term $a_2$}
From the expression for $\langle E_{k'}|f_k^{(1)}\rangle$, Eq.\eqref{eq:f1}, we obtain the second order correction
\begin{equation}
a_2=\langle f_k^{(1)}|f_{k}^{(1)}\rangle=\sum_{k'\not= k} |\langle E_{k'}|f_{k}^{(1)}\rangle|^2=\frac{1}{4}\sum_{k'\not=k} \frac{|\langle E_{k'}|\hat{K}|E_k\rangle|^2}{\sin^2\left(\frac{\tau(E_{k'}-E_k)}{2}\right)}.
\label{eq:a2exp}
\end{equation}
The numerators in the terms of the previous sum depend on the squared norm of matrix elements of the kick operator. For kick operators with well defined classical limit, these matrix elements scale linearly with $J$ for large enough $J$ (see  Fig. \ref{fig:3}(e) in main text) 
$$
|\langle E_{k'}|\hat{K}|E_k\rangle|\approx C_{k',k} J.
$$
As a consequence, the numerators give $a_2$ a quadratic dependence on $J$. If the state is non-resonant, the denominators evaluate to finite numbers, causing the whole term to scale quadratically with $J$, that is, $a_{2} \propto J^2$ for non-resonant states.

\begin{figure}
    \centering
    \includegraphics[width=1.0\linewidth]{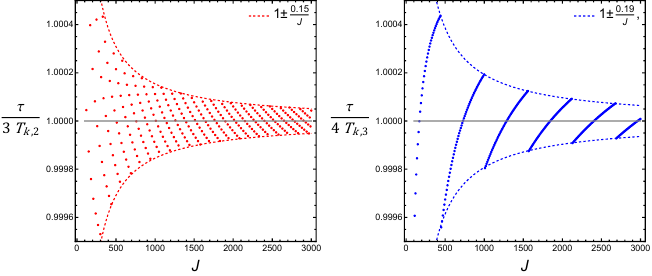}
    \caption{ Ratio  $\frac{\tau}{n T_{k,m}}$ that is closest to  satisfying  the  resonant condition $\frac{\tau}{n T_{k,m}}=1$ as a function of $J$ for fixed kick period $\tau=8$. Left panel is for resonance $m\!:\!n=2\!:\!3$ and  right panel  for resonance $m\!:\!n=3\!:\!4$. Dashed lines are fits to the decaying oscillations around $1$ of the ratios    $\frac{\tau}{n T_{k,m}}$. }
    \label{fig:tauT}
\end{figure}

However, if the state $|E_k\rangle$ is close to a resonance $m{:}n$, then for the term $k'=k+m$, we have $\tau (E_{k+m}-E_k)/2=n\pi+\delta$, where $\delta$ approaches zero as $\delta=\delta_{m:n}/J$ (see  Fig. \ref{fig:3}(d)  in main text and Fig.\ref{fig:tauT}). As a consequence, for this term $k'=k+m$, the denominator becomes small $$\sin\left(\frac{\tau(E_{k+m}-E_k)}{2}\right)\approx \frac{\delta_{m:n}}{J},$$ and, consequently, the term with this small denominator in the sum in Eq.\eqref{eq:a2exp} acquire an additional $J^2$ scaling, leading to $a_{2} \propto J^4 $ for close-to-resonance states.   

\begin{figure}
     \centering     
     \includegraphics[width=0.92\textwidth]{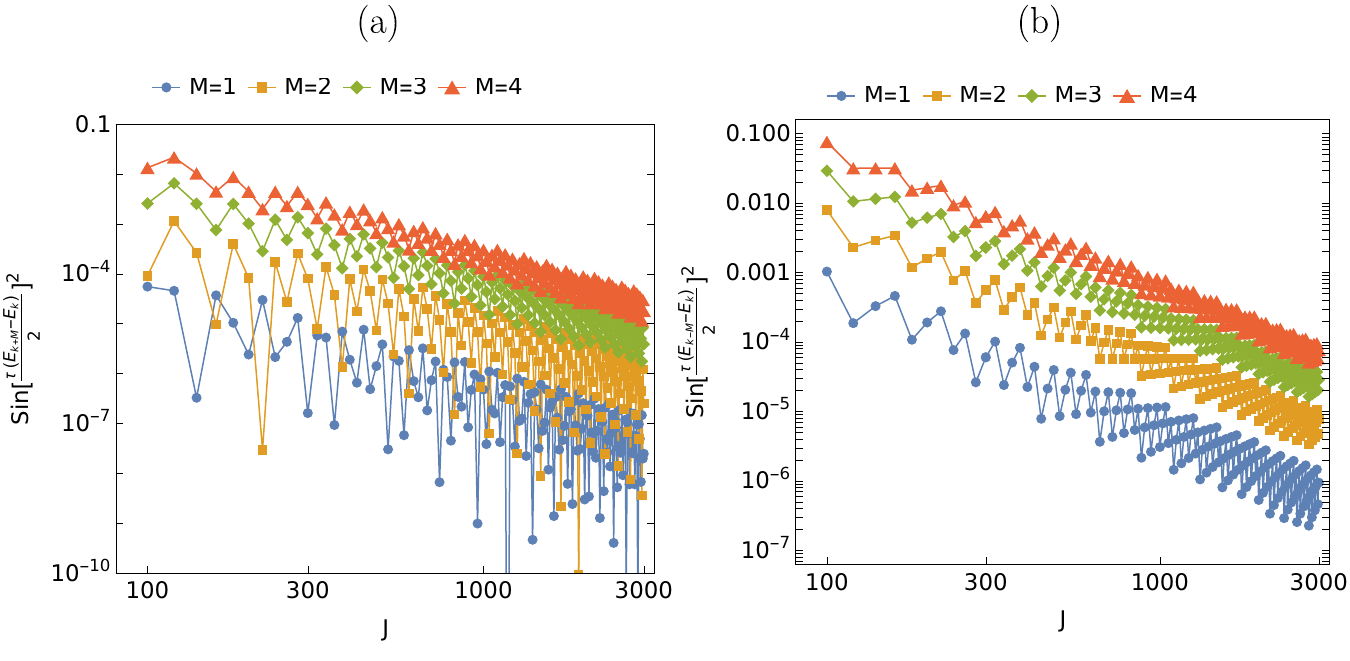}
     \caption{ (a) Log-log  plot of  $\sin^2\left(\tau \frac{E_{k+M}-E_{k}}{2}\right)$ as a function of $J$ for $M=1,2,3$ and $4$ for resonance $m{:}n=1:1$, where $E_k$ is the eigenvalue, for each $J$, that most closely satisfies the $1{:}1$ resonant condition, $\tau(E_{k+1}-E_k)/2=\pi$ for $\tau=8$. 
     (b) Similar to (a), but  for $\sin^2\left(\tau \frac{E_{k-M}-E_{K}}{2}\right)$ vs $J$.
    }.
     \label{fig:nextm1}
 \end{figure}

\begin{figure}
     \centering  
     \includegraphics[width=0.99\textwidth]{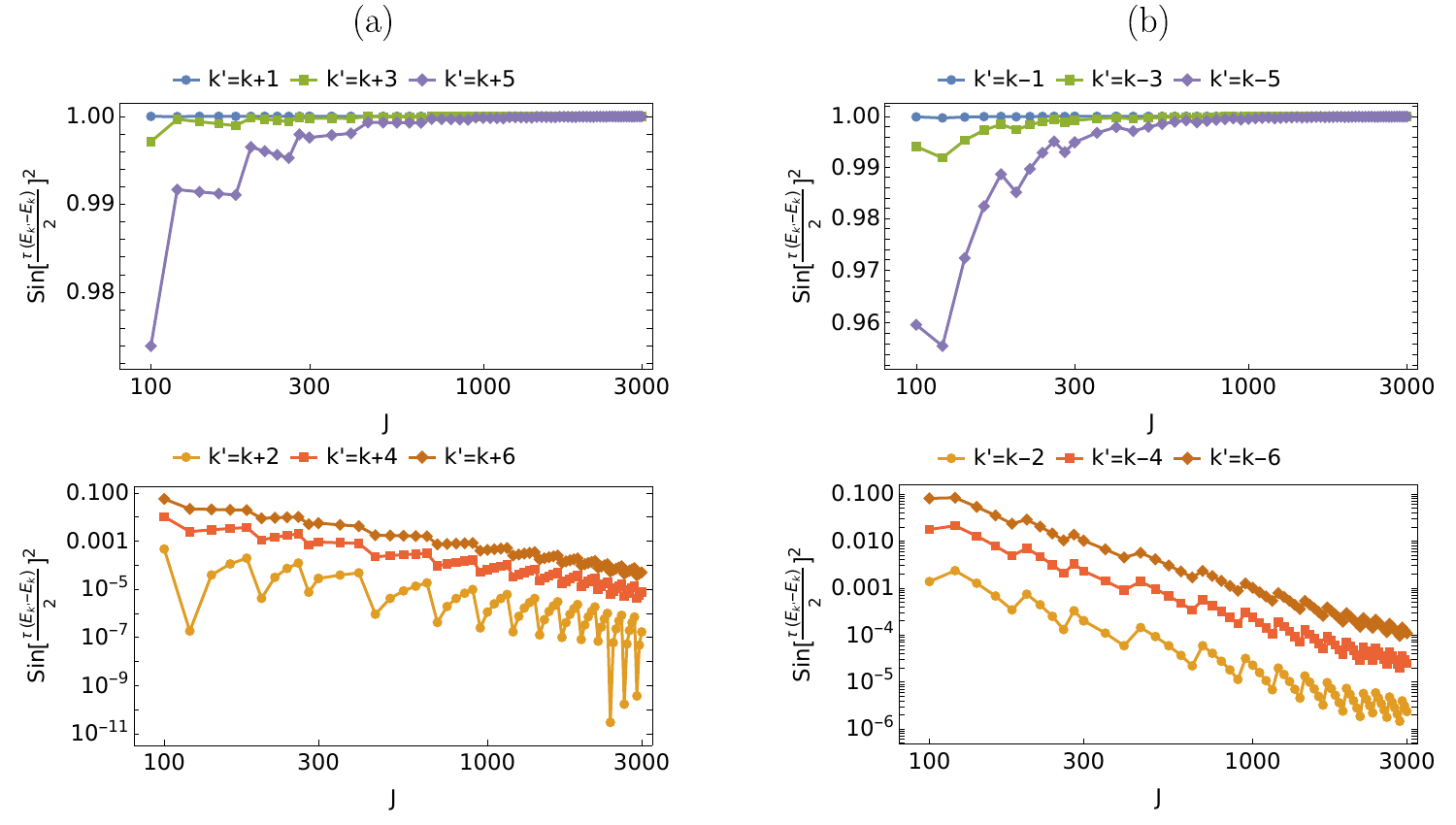}
     \caption{ Log-log plots  of  $\sin^2\left(\tau \frac{E_{k'}-E_{k}}{2}\right)$ as a function of $J$ for resonance $m{:}n=2:3$,  where $E_k$ is the eigenvalue, for each $J$, that most closely satisfies the $2{:}3$ resonant condition, $\tau(E_{k+2}-E_k)/2=3\pi$ for $\tau=8$. In panel (a) $k'=k+1,k+3$ and $k+5$ are non-resonant terms and they  are  close $1$ for all $J$, whereas  $k'=k+2,k+4,k+6$ are close-to-resonance terms that go to zero as $J$ increases. Panel (b) is the same for non-resonant ($k'=k-1,k-3$ and $k-5$)  and  close-to-resonance ($k'=k-2,k-4,k-6$) terms.   
     }.
     \label{fig:nextm2}
 \end{figure}

\begin{figure}
    \centering
    \includegraphics[width=0.9
    \linewidth]{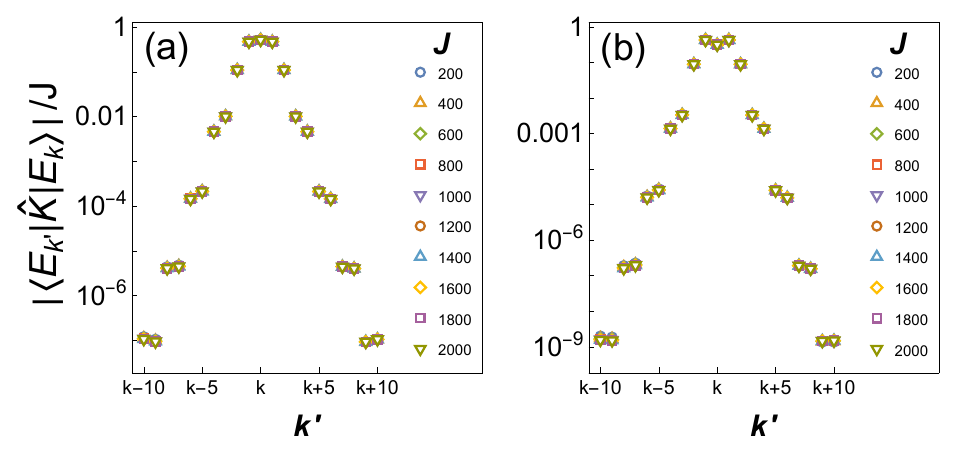}
    \caption{Log-linear plots of the magnitude  of the  kick-operator matrix elements divided by $J$,   $|\langle E_{k'}|\hat{K}|E_k\rangle|/J$,  where $|E_k\rangle$ is   the $\hat{H}_0$ eigenstate that comes closest to  satisfying the quantum resonant condition $m{:}n=1{:}1$ [panel (a)] and  $m{:}n=2{:}3$ [panel (b)]. Diagonal and first ten neighbor states, $|E_{k'}\rangle$,  above and below of  $|E_k\rangle$ are plotted. Different values of $J$, indicated in the panels, were used.      }
    \label{fig:matrixelem}
\end{figure}

Similar small denominators  appear in the   terms of the sum when   $k'=k+M m$ for $M$ integer and  $M\geq 2$. For these terms we have   $$\sin\left(\frac{\tau(E_{k+ Mm}-E_k)}{2}\right)\approx \sin\left(\frac{M \tau(E_{k+ m}-E_k)}{ 2}\right)\approx \frac{M \delta_{m:n}}{J}. $$ The dependence on $J$ of  the square of these terms is shown in  Figs.\ref{fig:nextm1}(a) and \ref{fig:nextm2}(a), for resonances $m{:}n=1{:}1$ and $2{:}3$, respectively. Small denominators appear also for terms  $k'=k- Mm$, for $M$ integer and  $M\geq 1$. In this case, $\sin\left(\frac{\tau(E_{k- Mm}-E_k)}{2}\right)\approx M \tilde\delta_{m:n}/J $,  with $\tilde{\delta}_{m:n}$ slightly  different from $\delta_{m:n}$. The square of these terms as a function of $J$ is shown in  Figs.\ref{fig:nextm1}(b) and \ref{fig:nextm2}(b), for resonances $m{:}n=1{:}1$ and $2{:}3$, respectively.   
However, all  the  terms in the sum with these latter denominators are smaller than the one closest to the resonance, not only by the factor $1/M^2$ 
but also because, for the kick operator considered here, the matrix elements are smaller for more distant states $| \langle E_{k\pm M m}|\hat{K}|E_k\rangle|<| \langle E_{k+m}|\hat{K}|E_k\rangle|$, see Fig.\ref{fig:matrixelem}. %
For resonances $m{:}n=1{:}n$, the previous terms exhaust all the possible terms in the sum of Eq.\eqref{eq:a2exp} ($k'=k\pm 1 M$), but for other resonances,  $m\geq2$, there are additional sum terms, that have no small denominators (see both panels of Fig.\ref{fig:nextm2}).  These remaining sum terms    scale only quadratically with $J$.    
In summary, for states close to resonance, the  coefficient $a_2$ has terms that scale as $J^4$ and other non-resonant terms that scale as $J^2$
$$
a_2=a_{2,\text{NR}}J^2+a_{2,\text{CR}} J^4 \xrightarrow[]{J\rightarrow\infty} a_{2,\text{CR}} J^4, 
$$
where  the quartic term dominates the quadratic one for large enough $J$ (in the case of resonances with $m=1$, the quadratic term is absent and the quartic term is dominant for all $J$). Meanwhile, for non-resonant states only  the quadratic term is present
$$
a_2=a_{2,\text{NR}}J^2.
$$

 \subsection{Third order term $a_3$}
Similarly, for higher order corrections to $\mathcal{F}_{k,\text{max}}$ (Eq.\eqref{eq:fmax4}), we obtain  the same splitting  between non-resonant and close-to-resonant terms in the sum, where the latter ones appear only for states close to  resonance. The scaling of $J$ in the third order term is discussed in the following.  From Eqs.\eqref{eq:f1} and \eqref{eq:f2}, the  term $a_3$ of the perturbative series for $\mathcal{F}_{k,\text{max}}$ can be written as
\begin{eqnarray}
   a_3=2\text{Re}\left[ \langle f_{k}^{(1)}|f_k^{(2)}\rangle\right]
  &=& \frac{1}{4}\sum_{k'\not=k}\sum_{k''\not=k,k'} \frac{\cos\left(\frac{\tau(E_k-E_{k''})}{2}\right) \langle E_k|\hat{K}|E_{k'}\rangle\langle E_{k'}|\hat{K}|E_{k''}\rangle\langle E_{k''}|\hat{K}|E_{k}\rangle}{\sin\left(\frac{\tau(E_k-E_{k''})}{2}\right) \sin^2\left(\frac{\tau(E_k-E_{k'})}{2}\right)}\nonumber\\
    &+&
    \frac{1}{4}\sum_{k'\not=k}\frac{\cos\left(\frac{\tau(E_k-E_{k'})}{2}\right)}{\sin^3\left(\frac{\tau(E_k-E_{k'})}{2}\right)}\abs{\langle E_k|\hat{K}|E_{k'}\rangle}^2\left(\langle E_{k'}|\hat{K}|E_{k'}\rangle-\langle E_{k}|\hat{K}|E_{k}\rangle\right)\nonumber
   \\
    &+&\frac{1}{8}\sum_{k'\not=k} \frac{\langle E_k|\hat{K}^2|E_{k'}\rangle\langle E_{k'}|\hat{K}|E_{k}\rangle-\langle E_k|\hat{K}|E_{k'}\rangle\langle E_{k'}|\hat{K}^2|E_{k}\rangle}{\sin^2\left(\frac{\tau(E_k-E_{k'})}{2}\right)}.
\end{eqnarray}
In the  second sum of the previous expression, we obtain a factor with the difference of diagonal matrix elements  $\hat{K}_{k'k'}-\hat{K}_{kk}$. Each matrix element scales linearly with $J$ but their difference tends to a constant for $J\rightarrow\infty$ (see Fig.\ref{fig:difK}), therefore the  numerators  in the terms of this second sum scale only quadratically  with $J$. Consequently, for non-resonant states this  sum scales quadratically with $J$; whereas for states close to a resonance, the  sum contains non-resonant terms scaling quadratically and other close-to-resonant terms scaling as $J^5$ (where a factor of $J^3$ arises from small denominators).    In the third sum, if the matrix elements of $\hat{K}$ are real (as is the case considered in this work), the numerators vanish. The main contribution to $a_3$ in the limit $J\rightarrow \infty$ comes from the first sum, which scales as $J^3$ for non-resonant states, and as $J^6$ for close-to-resonance states (which includes terms scaling as $J^3$ and resonant terms with small denominators scaling as $J^6$).  Thus, 
\begin{align}
&a_3= a_{3,\text{NR},1}J^2+a_{3,\text{NR},2}J^3 &\text{\ \ for non-resonant states}\\
&a_3= a_{3,\text{NR},1}J^2+a_{3,\text{NR},2}J^3+a_{3,\text{CR},1}J^5+a_{3,\text{CR},2}J^6 & \text{\ \ for states close to  resonance}
\end{align}
In the limit $J\rightarrow\infty$, we obtain $a_3\approx a_{3,\text{NR},2} J^3$ and $a_3\approx a_{3,\text{CR},2}J^6$ for non-resonant and close-to-resonance  states, respectively.

\subsection{Scaling of $\epsilon_\text{max}$ from the perturbative series}

\begin{figure}
    \centering
    \includegraphics[width=0.9\linewidth]{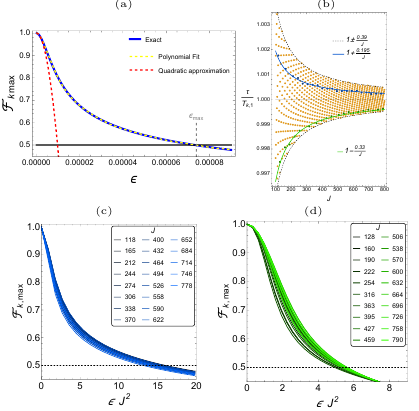}   
    \caption{(a) Continuous blue line shows the exact numerical  $\mathcal{F}_{k,\text{max}}$ as a function of $\epsilon$ for  the state closest to the resonance  $1{:}1$,  for $\tau=8$, and $J=500$. Red dashed line is the perturbative approximation until second order in $\epsilon$. The condition $\mathcal{F}_{k,\text{max}}=1/2$ (horizontal black line) determines $\epsilon_\text{max}$ (vertical dashed line). Yellow dashed line is a polynomial of degree $10$, $p(\epsilon)=1-\sum_{p=2}^{10} b_k \epsilon^p$, fitted to  the exact numerical data. (b) Orange dots depict the  kick period over the quantum period that is closest to satisfying the $m{:}n=1{:}1$ quantum resonant condition $\frac{\tau}{T_{k,1}}=1$ as a function of $J$. Coloured  dots indicate  values of $J$ for which the ratio $\frac{\tau}{T_{k,1}}$ approaches $1$ without oscillating, from above (blue dots) and from below (green dots). These values of $J$ are used in panel (c) and (d), respectively. (c) $\mathcal{F}_{k,\text{max}}$ as a function of $z=\epsilon J^2$ for the values of $J$ corresponding to the blue dots of panel (b). (d) Same as panel (c) but for the green dots of panel (b).
    }
   \label{fig:fmax400}
\end{figure}
\begin{figure}
    \centering
    \includegraphics[width=0.8\linewidth]{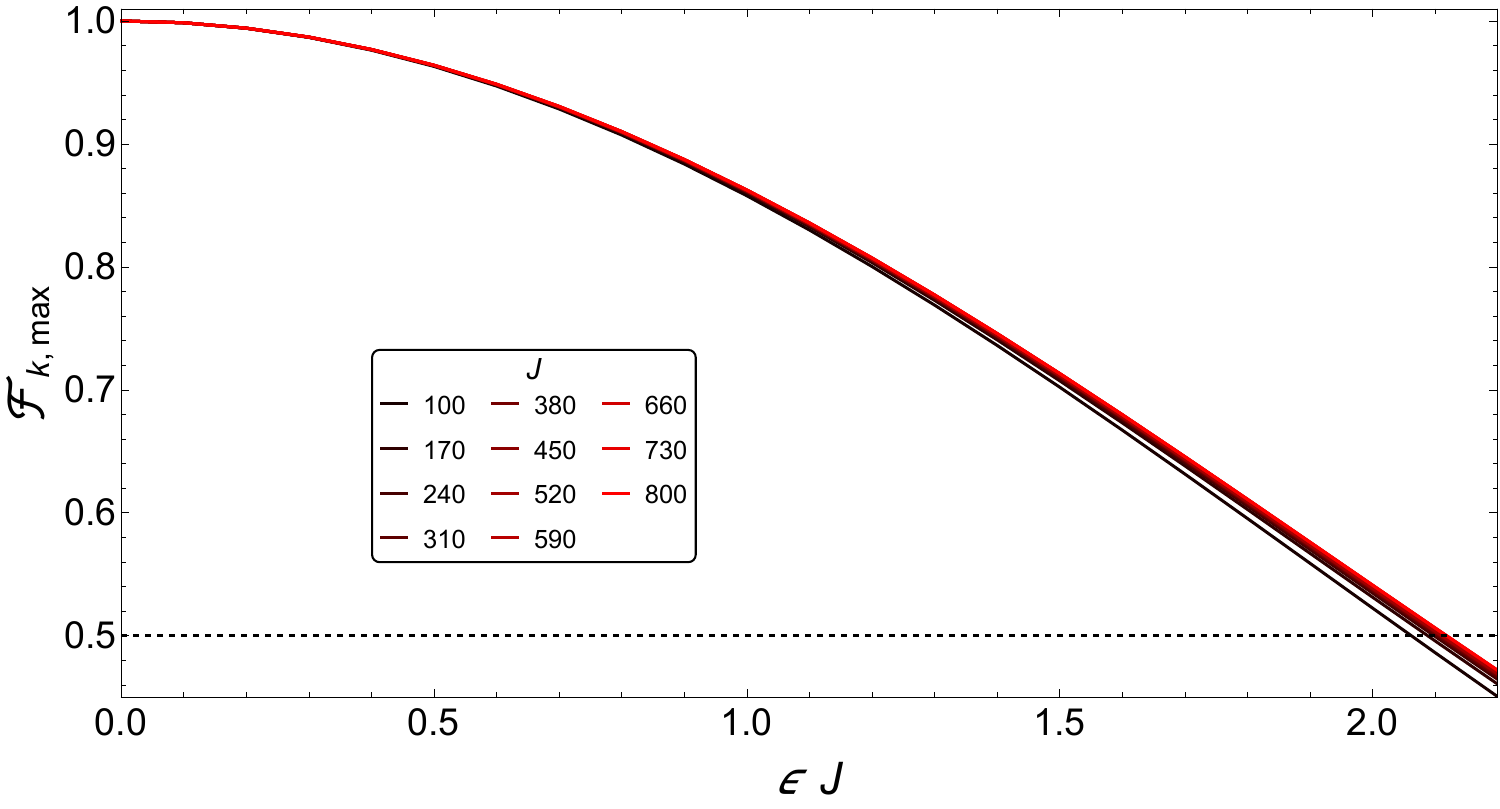}
    \caption{Numerically exact $\mathcal{F}_{k,\max}$ for a non-resonant state as a function of $z=\epsilon J$ for the indicated values of $J$. All the curves cluster around one common curve. For each value of $J$, we consider the state $|E_k\rangle$  with energy closest to $E_k\approx E/J=0$, where no resonances are present.}
    \label{fig:FmaxScNR}
\end{figure}

A generalization of the dependence on $J$ that we found for the second and third order perturbative coefficients of $\mathcal{F}_{k,\text{max}}$  explains the scaling obtained for $\epsilon_{\text{max}}$ in the main text, that is, $\epsilon_{\text{max}}^{\text{NR}} \propto 1/J$ and $\epsilon_{\text{max}}^{\text{CR}} \propto 1/J^2$. In previous subsection, we found that for large enough $J$, the coefficients for non-resonant states scale as  $a_2=a_{2,\text{NR}}J^2$ and $a_3=a_{3,\text{NR}}J^3$, whereas  for close-to-resonance states $a_2=a_{2,\text{CR}}J^4$ and $a_3=a_{3,\text{CR}}J^6$. The generalization of these scalings for higher-order coefficients, 
$a_{\text{p}}= a_{\text{p},\text{NR}}J^\text{p}$ and $a_\text{p}=a_{\text{p},\text{CR}}J^{2 \text{p}},$ 
imply that the maximum fidelity is 
$$
\mathcal{F}_{k,\text{max}}=1-\sum_{\text{p}=2}^{\text{p}_o}a_\text{p} \epsilon^{\text{p}}= 1- \sum_{\text{p}=2}^{\text{p}_o}a_{p,\text{NR}}(\epsilon J)^\text{p}
$$
$$
\mathcal{F}_{k,\text{max}}= 1-\sum_{\text{p}=2}^{\text{p}_o}a_\text{p} \epsilon^{\text{p}}=1- \sum_{\text{p}=2}^{\text{p}_o}a_{\text{p},\text{CR}}(\epsilon J^2)^\text{p},
$$
for resonant and close-to-resonance cases, respectively, where $\text{p}_o$ is the order of the perturbative series that is needed to accurately describe $\mathcal{F}_{k,\text{max}}$ in the interval $\epsilon\in[0,\epsilon_{k,\text{max}}]$, where $\mathcal{F}_{k,\text{max}}\geq 1/2$. Regardless  of $\text{p}_o$, we have that $\mathcal{F}_{k,\text{max}}$ is a function of variable $z=\epsilon J$ for non-resonant states and a function of variable $z=\epsilon J^2$ for states close to resonance. This differentiated dependence of $\mathcal{F}_{k,\text{max}}$ between resonant and close-to-resonance states allows us to explain the numerically observed scalings of $\epsilon_\text{max}$, as explained in the main text and we discuss it in detail in the next paragraph.   

In  Fig. \ref{fig:fmax400}(a), we show  that the  perturbative series of $\mathcal{F}_{k,\text{max}}$ up to $\order{\epsilon^2}$ becomes inaccurate well
before $\mathcal{F}_{k,\text{max}}$ attains the value $1/2$.  In the figure,  we compare  the numerically exact and quadratic approximation of  $\mathcal{F}_{k, \text{max}}$ as a function of $\epsilon$, for $J=500$ in the case of the state close to the resonance $m{:}n=1{:}1$.  The quadratic approximation accurately represents the numerical $\mathcal{F}_{k, \text{max}}$ only  for a very small interval of $\epsilon$ close to zero. Higher order corrections are needed (at least up to order $\text{p}_o=10$, yellow dashed-line in the same figure) to accurately describe the numerical results in the whole range $\epsilon\in[0,\epsilon_{k,\text{max}}]$.
Even if the $\order{\epsilon^2}$ is unable to describe $\mathcal{F}_{k,\text{max}}$, our analysis and discussion about  the perturbative series indicate that $F_{k,\text{max}}=1-a_{2,\text{CR}} (\epsilon J)^2-...-a_{10,\text{CR}} (\epsilon J)^{10}$. Then, to determine  $\epsilon_{k,\text{max}}$ we have to find  the roots of the degree-10 polynomial equation
$$1-a_{2,\text{CR}} (\epsilon J^2)^2-...-a_{10,\text{CR}} (\epsilon J^2)^{10}=1-a_{2,\text{CR}} z^2-...-a_{10,\text{CR}} z^{10}=\frac{1}{2},$$ where $z=\epsilon J^2$.  From the ten roots of the previous equation, one of them determines $\epsilon_{k,\text{max}}$. We call it $z_{\text{CR}}$.  From this root, we have  $\epsilon_{k,\text{max}} J^2=z_{\text{CR}}$, i.e.,  $\epsilon_{k,\text{max}}=z_{CR}/J^2$, which is the scaling obtained numerically.  
Similarly, for non-resonant states, we  solve a polynomial equation, but this time in terms  of variable $z=\epsilon J$. If we call $z_{\text{NR}}$ the relevant solution to the polynomial equation, we obtain $\epsilon_{k,\text{max}}=z_{\text{NR}}/J$, which is, again, the scaling obtained numerically.  

Since the perturbative coefficients become very complicated for  orders higher than $\order{\epsilon^3}$, it is a very hard task to verify from  perturbative calculations that, for $\text{p}>3$, $a_{\text{p}}=a_{2,\text{NR}}J^\text{p}$ and $a_{\text{p}}=a_{2,\text{CR}}J^{2\text{p}}$. However, it is possible to verify numerically that, indeed $F_{k,\text{max}}$ is a function of $z=\epsilon J$ for non-resonant states and of $z=\epsilon J^2$ for close-to-resonance states in the interval of interest where $F_{k,\text{max}}\geq 1/2$. This is done in Fig.\ref{fig:FmaxScNR} for a non-resonant state. The figure shows $F_{k, \text{max}}$ as a function of $z=\epsilon J$ for different values of $J$. It is clearly observed that all the curves cluster  around a common curve, which indicates that effectively, $F_{k,\max}$ is a function of $z=\epsilon J$. For close-to-resonance states a similar collapse of curves  is obtained, however, in this case  a more careful procedure has to be done to observe this. We already know  that for  states close to resonance, the ratio $\frac{\tau}{n T_{k,m}}$  oscillates around $1$, with an amplitude that decays as $1/J$.  To eliminate the effect of the oscillations, we select different values of $J$ for which the ratio $\frac{\tau}{n T_{k,m}}$  decays without oscillating. In Fig.~\ref{fig:fmax400}(b), we indicate in blue and green  two sets of $J$ values for which the ratio $\frac{\tau}{ T_{k,1}}$ (we consider the resonance $m{:}n=1{:}1$) decays as $1/J$ without oscillating. One of these sets has ratios  $\frac{\tau}{ T_{k,1}}>1$ (blue set) and the other has ratios $\frac{\tau}{ T_{k,1}}<1$ (green set). In panels (c) and (d) of Fig.~\ref{fig:fmax400}, we show the numerically exact evaluation of $\mathcal{F}_{k,\text{max}}$ as a function of $z=\epsilon J^2$, for all the values of $J$ in each set. We observe that, indeed, all the curves cluster  around the same common curve, substantiating our claim that $F_{k,\text{max}}$ is a function of $z=\epsilon J^2$ for close-to-resonance states.

\subsection{Dominance of the largest power of $J$ in the coefficients of the $\mathcal{F}_{k,\text{max}}$ perturbative series}\label{subsec:Dominance}

\begin{figure}
    \centering
    \includegraphics[width=0.99\linewidth]{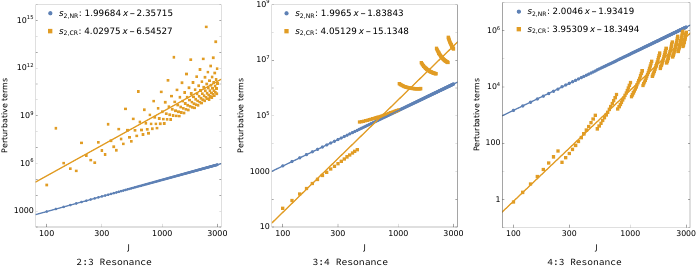}
    \caption{Log-log plots showing the close-to-resonant (\(s_{2,\text{CR}}\), orange dots) and non-resonant (\(s_{2,\text{NR}}\), blue dots) contributions to the second-order correction \(a_2\) to \(\mathcal{F}_{k,\text{max}}\), as described in Eq. \eqref{eq:a2spilt}, as a function of \(J\) for the resonances \(2{:}3\), \(3{:}4\), and \(4{:}3\). Linear fits in ln-ln scale to the data points are indicated. Within the range shown, the close-to-resonant contribution \(s_{2,\text{CR}}\) is dominant over the non-resonant contribution \(s_{2,\text{NR}}\) only for the \(2{:}3\) resonance. From the fits shown in the panels: $s_{2,\text{NR}}\approx0.09469 J^2$ and $s_{2,\text{CR}}\approx 0.001437 J^4$ for resonance $2{:}3$; $s_{2,\text{NR}}\approx 0.1591 J^2$ and $s_{2,\text{CR}}\approx 2.673\times10^{-7} J^4$ for resonance $3{:}4$; and $s_{2,\text{NR}}\approx 0.1445 J^2$ and $s_{2,\text{CR}}\approx 1.074\times10^{-8} J^4$ for resonance $4{:}3$.
    }
    \label{fig:terms}
\end{figure}

In order to obtain the universal scalings of  $\epsilon_\text{max}$ in the limit $J\rightarrow\infty$, it is necessary for the terms with the largest power of $J$ in  the coefficients of the perturbative series of $\mathcal{F}_{k,\text{max}}$ to dominate. 
Here, we analyze how large $J$ must be for this condition to be met in the close-to-resonance case,
and discuss whether the numerically available values of $J$ are large enough for the resonances $1\!:\!1$, $2\!:\!3$ and $3\!:\!4$ in the case of the Hamiltonian we consider  in the main text.

As an example, we consider the close to $m{:}n = 2{:}3$ resonance case. First, we split the sum  in Eq.(\ref{eq:a2exp}) as 
\begin{equation}
a_2=s_{2,\text{CR}}+s_{2,\text{NR}},
\label{eq:a2spilt}
\end{equation} where $s_{2,\text{CR}}$ contains close-to-resonance terms and $s_{2,\text{NR}}$ contains the remaining ones.  The close-to-resonance terms are those that approximately satisfy $\tau(E_{k'}-E_k)/2\approx 3 \pi$, i.e., those with $k'=k\pm 2M$ where $M$ is a strictly-positive integer. These terms give   
$$
s_{2,\text{CR}}=\frac{1}{4}\sum_{M=\pm 1,\pm 2,...} \frac{|\langle E_{k+ 2M}|\hat{K}|E_k\rangle|^2}{\sin^2\left(\frac{\tau(E_{k+2M}-E_k)}{2}\right)}. 
$$
From all these terms, the most important  is the one closest to the resonant condition, $M=1$. Thus,

$$
s_{2,\text{CR}}\approx \frac{|\langle E_{k+2}|\hat{K}|E_k\rangle|^2}{4\sin^2\left(\frac{\tau(E_{k+2}-E_k)}{2}\right)}\approx
\frac{|\langle E_{k+2}|\hat{K}|E_k\rangle|^2 J^2}{4 \delta_{2:3}^2}\approx
  \frac{C_{k+2,k}^2}{4\delta_{2:3}^2} J^4=a_{2,\text{CR}}J^4,
$$
where we have used  that $\tau(E_{k+2}-E_k)/2\approx 3 \pi + \delta_{2:3}/J$, and expressed  $|\langle E_{k+2}|\hat{K}|E_k\rangle|= C_{k+2,k} J$.

From all the remaining (non-resonant) terms, the most important  are the ones involving first neighbors ($k'=k\pm 1$)
$$
s_{2,\text{NR}}\approx \frac{|\langle E_{k+1}|\hat{K}|E_k\rangle|^2}{4\sin^2\left(\frac{\tau(E_{k+1}-E_k)}{2}\right)}+ \frac{|\langle E_{k-1}|\hat{K}|E_k\rangle|^2}{4\sin^2\left(\frac{\tau(E_{k-1}-E_k)}{2}\right)}\approx
$$
$$
\frac{|\langle E_{k+1}|\hat{K}|E_k\rangle|^2}{4}+ \frac{|\langle E_{k-1}|\hat{K}|E_k\rangle|^2}{4}\approx \frac{|\langle E_{k+1}|\hat{K}|E_k\rangle|^2}{2}\approx \frac{C_{k+1,k}^2}{2} J^2=a_{2,\text{NR}}J^2 ,
$$
where we have used $\tau(E_{k\pm1}-E_k)/2\approx \tau(E_{k\pm 2}-E_k)/4\approx \pm 3\pi/2$ [using aforementioned condition for $2{:}3$ resonance and $E_{k\pm 2}-E_k\approx 2(E_{k\pm 1}-E_k)$], assumed that $|\langle E_{k+1}|\hat{K}|E_k\rangle|\approx |\langle E_{k-1}|\hat{K}|E_k\rangle|$, and  expressed $|\langle E_{k+1}|\hat{K}|E_k\rangle|=C_{k+1,k} J$. From the previous expressions, the condition of dominance of the quartic term, $a_{2,\text{NR}}J^2 \ll a_{2,\text{CR}}J^4$,  can be  written as $$\frac{2 |\delta_{2:3}|C_{k+1,k}}{C_{k+2,k}}\ll J.$$ 
For the case studied in the main text, and using a tolerance of $10^{-3}$ for the ratio between non-resonant and resonant terms ($a_{2,\text{NR}}J^2=10^{-3} a_{2,\text{CR}}J^4$, see numerical estimates of $a_{2,\text{NR}}$ and $a_{2,\text{CR}} $ in Fig.~\ref{fig:terms}), we 
obtain $J>256$ for resonance $2\!\!:\!\!3$, which are available system sizes  for our numerical methods. 

 \begin{figure}
     \centering
     \includegraphics[width=0.6\linewidth]{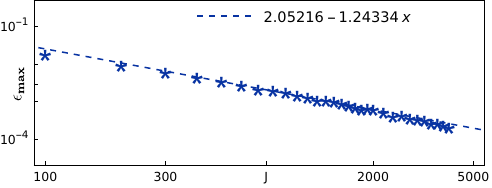}
     \caption{Log-log plot  showing the scaling of $\epsilon_{\text{max}}^{\text{CR}}$ with $J$ for $3{:}4$ resonance. For the range of $J$ values considered, we get $\epsilon_{\text{max}}^{\text{CR}} \propto J^{-1.24}$, far from $J^{-2}$. As explained in  subsection~\ref{subsec:Dominance}, $J> 2.4\times 10^4$ needs to be considered to obtain a scaling of $J^{-2}$. Data available at \cite{Raw_data_doi} in the main text.}
     \label{fig:SM34scaling}
 \end{figure}

Repetition of the same analysis for resonance $3{:}4$ results in the condition  $$\frac{2 |\delta_{3:4}| C_{k+1,k}}{C_{k+3,k}}\ll  J,$$
where $ |\langle E_{k+3}|\hat{K}|E_k\rangle|\approx C_{k+3,k} J$, and $\tau(E_{k+3}-E_k)/2\approx 4\pi+\delta_{3:4}/J$.
Considering the same tolerance  as before  ($a_{2,\text{NR}}J^2=10^{-3} a_{2,\text{CR}}J^4$, see Fig.~\ref{fig:terms} for  numerical estimations of $a_{2,\text{NR}}$ and $a_{2,\text{CR}} $),
we now obtain  $J>24 393$, which is beyond the values of $J$ we consider in this work. As a consequence, the dominant scaling of $\epsilon_\text{max}$ for $J\rightarrow \infty$ is achieved for moderate system sizes for low-order resonances $1{:}1$ and $2{:}3$, while for higher-order resonances ($3{:}4$ and others), we would have to consider larger values of $J$. This is illustrated in Fig.\ref{fig:terms}, where the close-to-resonant  $s_{2,\text{CR}}$  and non-resonant $s_{2,\text{NR}}$ parts of  $a_2$ are plotted as a function of $J $  in the range $J\in[100,3000]$ for three resonances: $2{:}3$, $3{:}4$ and $4{:}3$. In the case of resonance $2{:}3$, the close-to-resonance part $s_{2,\text{CR}}$ is much larger than $s_{2,\text{NR}}$, while for the other resonances the close-to-resonant part $s_{2,\text{CR}}$ is barely larger  or even smaller than the non-resonant part $s_{2,\text{NR}}$. From this, it is concluded that the universal scaling for $\epsilon_\text{max}$ is available for moderate sizes of $J\in[100,4000]$  in the case of resonance $1{:}1$ (for this resonance, the non-resonant contribution vanishes, $s_{2,\text{NR}}=0$) and resonance $2{:}3$. For the resonance $3{:}4$, much larger values of $J$ than those used in this work are needed, such as $J\in[24000,2.4\times 10^5]$. In Fig.\ref{fig:SM34scaling}, we show the $\epsilon_\text{max}$ scaling analysis for this resonance $m{:}n=3{:}4$ in the range $J\in[100,4000]$,  similar to those shown in  Fig.\ref{fig:3}(b) of the main text  for  resonances  $m{:}n=1{:}1$ and $2{:}3$. Unlike those resonances, in this case we obtain $\epsilon_\text{max}\propto J^{1.243}$, very far from the expected scaling for $J\rightarrow\infty$, $\epsilon_\text{max}\propto J^2$.

\section{Perturbative series for $\mathcal{F}_\text{max}$: exact resonance case}
\label{sec:ApPFER}
In this section, we present the details to obtain perturbatively  $\mathcal{F}_\text{max}$ in the case of an exact resonance involving states $|E_{k}\rangle$ and $|E_{k+m}\rangle$. We analyze the  dependence on $J$ of its different terms and from that we  analytically obtain the scaling of $\epsilon_\text{max}$ with the system size, $J$.

The normalized Floquet eigenstate in degenerate perturbation theory, which applies for exact resonant states, is 
\begin{align}
|f_{a}\rangle=\frac{1}{\sqrt{1+\epsilon^2 \langle f_a^{(1)}|f_{a}^{(1)}\rangle+ \order{\epsilon^3} }}\left( |f_a^{(0)}\rangle+ \epsilon |f_a^{(1)}\rangle+ \epsilon^2 |f_{a}^{(2)}\rangle +\mathcal{O}(\epsilon^3)\right)=\\
\left(1-\frac{1}{2} \langle f_a^{(1)}|f_{a}^{(1)}\rangle \epsilon^2 \right)|f_a^{(0)}\rangle+ \epsilon |f_a^{(1)}\rangle+ \epsilon^2 |f_{a}^{(2)}\rangle +\mathcal{O}(\epsilon^3).
\end{align}

From this, the maximum overlap of the Floquet eigenstate  is obtained by consistently  expanding  in powers of $\epsilon$ the squared norm $|\langle E_{a}|f_{a}\rangle|^2$. The result up to $\order{\epsilon^2}$ is 
\begin{align} \mathcal{F}_{a,\text{max}}&=|\langle E_{a}|f_{a}\rangle|^2=\left(1-\epsilon^2\langle f_{a}^{(1)}|f_{a}^{(1)}\rangle\right)|\langle E_a|f_{a}^{(0)}\rangle|^2+2 \text{Re}\left[\langle E_{a}|f_{a}^{(0)}\rangle\langle E_a|f_{a}^{(1)}\rangle^*\right]\epsilon+ \label{Eq:fmaxdeg}\\
    & \left(|\langle E_a|f_a^{(1)}\rangle|^2 +2\text{Re}\left[\langle E_{a}|f_{a}^{(0)}\rangle\langle E_a|f_{a}^{(2)}\rangle^* \right]\right) \epsilon^2=\nonumber\\
&\left(1-\epsilon^2\sum_{k'\not=a,a'} |\langle E_{k'}|f_{a}^{(1)}\rangle|^2\rangle\right)|\langle E_a|f_{a}^{(0)}\rangle|^2+
2 \text{Re} \left[\langle E_{a}|f_{a}^{(0)}\rangle\langle f_{a'}^{(0)}|E_{a}\rangle \langle f_a^{(1)}|f_{a'}^{(0)}\rangle\right]\epsilon+ \nonumber\\
&\left[|\langle E_a|f_{a'}^{(0)}\rangle|^2-|\langle E_a|f_{a}^{(0)}\rangle|^2 \right]|\langle f_{a'}^{(0)}|f_{a}^{(1)}\rangle|^2 \epsilon^2+2\text{Re}\left[\langle E_{a}|f_{a}^{(0)}\rangle\langle E_a|f_{a}^{(2)}\rangle^* \right] \epsilon^2,\label{eq:Fmaxexp}
\end{align}
where $a,a'=k$ or $k+m$ ($a'\not=a$) and in the last equality we have used  
$
\langle E_a|f_a^{(1)}\rangle=\langle E_a|f_{a'}^{(0)}\rangle\langle 
 f_{a'}^{(0)}|f_{a}^{(1)}\rangle
$
and
$$
\langle f_{a}^{(1)}|f_{a}^{(1)}\rangle=\sum_{k'\not=a,a'} |\langle E_{k'}|f_a^{(1)}\rangle|^2+ |\langle f_{a'}^{(0)}|f_{a}^{(1)}\rangle|^2.$$
We will show that the maximal fidelity for large $J$ can be very well approximated by considering  only the first term of Eq.\eqref{eq:Fmaxexp},
\begin{equation}     \mathcal{F}_{a,\text{max}}=|\langle E_{a}|f_{a}\rangle|^2\approx\left(1-\epsilon^2\sum_{k'\not=a,a'} |\langle E_{k'}|f_a^{(1)}\rangle|^2\right)|\langle E_a|f_{a}^{(0)}\rangle|^2.
     \label{eq:fmaxapp}
\end{equation}

In order to show this, we now determine the dependence on $J$ of the different terms in the complete expression of $\mathcal{F}_{a,\text{max}}$, Eq.~\eqref{eq:Fmaxexp}. We begin by analyzing  the dependence on $J$ of the sum in the first term  of Eq.~\eqref{eq:Fmaxexp},
$$
\sum_{k'\not=a,a'} |\langle E_{k'}|f_a^{(1)}\rangle|^2=\sum_{k'\not=a,a'}\frac{|\langle E_{k'}|\hat{K}|f_{a}^{(0)}\rangle|^2}{4\sin^2\left(\frac{\tau(E_{k'}-E_a)}{2}\right)}.$$
Similar to  states close to resonance, this sum  scales as $J^4$.  This scaling  comes from the linear scaling of the matrix elements and from small denominators. In order to simplify the presentation, from now on, we assume that $a=k$ and $a'=k+m$. The dominant terms with small denominators in the sum are those coming  from  $k'=a-m$ and $k'=a+2m$, for which $\tau(E_{k+2m}-E_k)\approx 4\pi n+ \frac{\delta_{m:n}}{J}$ and $ \tau(E_{k-m}-E_k)\approx -2\pi n+ \frac{\delta_{m:n}}{J}$. From this, we obtain  
\begin{equation}\sum_{k'\not=a,a'} \frac{|\langle E_{k'}|\hat{K}|f_{a}^{(0)}\rangle|^2}{4\sin^2\left(\frac{\tau(E_{k'}-E_a)}{2}\right)}\approx \frac{|\langle E_{k-m}|\hat{K}|f_{k}^{(0)}\rangle|^2}{4 \delta_{m:n}^2}J^2 +\frac{|\langle E_{k+2m}|\hat{K}|f_{k}^{(0)}\rangle|^2}{4 \delta_{m:n}^2}J^2  =  D_o J^4.
\label{eq:SumScaling}
\end{equation}

\subsection{ Dependence on $J$ of $\langle E_{a}|f_a^{(0)}\rangle$ 
 and $\langle E_{a}|f_{a'}^{(0)}\rangle$ }
 
\begin{figure}
    \centering
        \includegraphics[width=\linewidth]{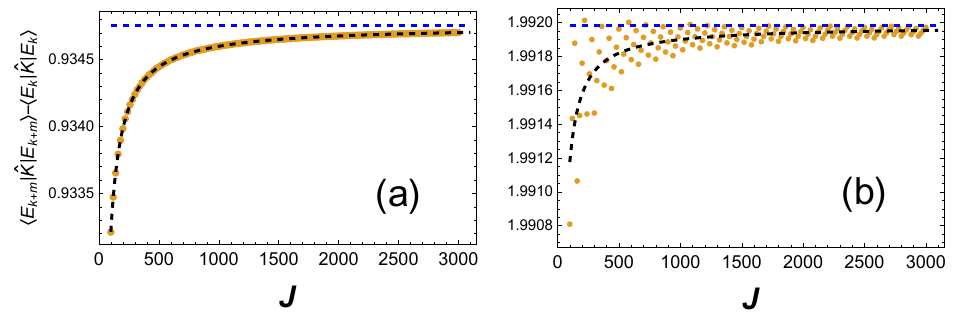}

    \caption{Orange dots depict $\left(\langle E_{k+m}|\hat{K}|E_{k+m}\rangle-\langle E_{k}|\hat{K}|E_{k}\rangle\right)$ vs $J$, where $|E_k\rangle$ is the $\hat{H}_0$ eigenstate  that comes closest to satisfying the quantum resonant condition $m{:}n$. Panel  (a) is  for resonance $m{:}n=1{:}1$ and (b) for resonance  $m{:}n=2{:}3$.  It can be observed that these differences,  for large $J$,   tend to a constant value, $\kappa_o$ (blue dashed lines). Black dashed lines are fits of the form $\kappa_o+ \frac{C}{J}$. 
    }
    \label{fig:difK}
\end{figure}

Now we analyze the dependence on $J$ of $\langle E_{a}|f_a^{(0)}\rangle$ and $\langle E_{a}|f_{a'}^{(0)}\rangle$. To do that, we have to diagonalize the kick operator in the degenerate subspace, Eq.~(\ref{eq:degem}). The scaling of its matrix elements is linear
 \begin{eqnarray}
      K_{k+m,k}=\langle E_{k+m}|\hat{K}|E_{k}\rangle&= \kappa J\\
     K_{k,k}=\langle E_{k}|\hat{K}|E_{k}\rangle&=& \kappa_1 J\\
      K_{k+m,k+m}=\langle E_{k+m}|\hat{K}|E_{k+m}\rangle&=& \kappa_2 J,   
 \end{eqnarray}
 where we have assumed that the non-diagonal term $K_{k+m,k}$ is real and without loss of generality   $K_{k+m,k}>0$, which implies that $\kappa>0$. 
The diagonal matrix elements scale linearly with $J$ but,  as shown in Figure \ref{fig:difK}, their difference  tends to be constant in the limit $J\rightarrow\infty$,  
$$
\lim_{J\rightarrow\infty} (\langle E_{k+m}|\hat{K}|E_{k+m}\rangle-\langle E_{k}|\hat{K}|E_{k}\rangle)=\kappa_o.
$$

From the previous behavior of the matrix elements, the matrix to be diagonalized for large $J$ is
\begin{eqnarray}
\left(\begin{array}{cc}
     K_{k,k}&K_{k,k+m}  \\
     K_{k+m,k}&K_{k+m,k+m} 
\end{array}\right)& = & K_{k,k}\left(\begin{array}{cc}
     1&0  \\
     0&1 
\end{array}\right)+\left(\begin{array}{cc}
     0&K_{k,k+m}  \\
     K_{k+m,k}&K_{k+m,k+m} - K_{k,k}
\end{array}\right)  \nonumber\\
&\xrightarrow[]{J\to \infty}& \kappa_1 J \left(\begin{array}{cc}
     1&0  \\
     0&1 
\end{array}\right)+ J\left(\begin{array}{cc}
     0&\kappa   \\
     \kappa  &\kappa_o/J
\end{array}\right) .
\end{eqnarray}
The eigenvalues and eigenvectors of this matrix, at leading order in  $1/J$, are 
\begin{eqnarray}
    \phi_{k}^{(1)}=(\kappa_1 - \kappa)J + \frac{\kappa_o}{2}+\mathcal{O}\left(\frac{1}{J}\right) \label{eq:quada} \\
    \phi_{k+m}^{(1)}=(\kappa_1 + \kappa)J+  \frac{\kappa_o}{2}+\mathcal{O}\left(\frac{1}{J}\right)\label{eq:quadap}
\end{eqnarray}
and 
\begin{eqnarray}
|f_{k}^{(0)}\rangle&=&\frac{1}{\sqrt{2}}\left(1+\frac{\kappa_o}{4 \kappa J}\right) |E_{k}\rangle-\frac{1}{\sqrt{2}}\left(1-\frac{\kappa_o}{4 \kappa J}\right) |E_{k+m}\rangle+ \mathcal{O}\left(\frac{1}{J^2}\right) \label{eq:fapA}\\
|f_{k+m}^{(0)}\rangle&=&\frac{1}{\sqrt{2}}\left( 1-\frac{\kappa_o}{4 \kappa J}\right) |E_{k}\rangle+\frac{1}{\sqrt{2}}\left(1+\frac{\kappa_o}{4 \kappa J}\right) |E_{k+m}\rangle +\mathcal{O}\left(\frac{1}{J^2}\right)\label{eq:fapB},
\end{eqnarray}
where, in order to label the states $|f_{a}^{(0)}\rangle$ and $\phi_a^{(1)}$  we have assumed that $\kappa_0>0$  (if $\kappa_0<0$, the labels simply have to be interchanged).   
From the previous equations, we obtain 
\begin{eqnarray}
    c_{k,1}=\langle E_k| f_{k}^{(0)}\rangle&=&\frac{1}{\sqrt{2}}\left(1+\frac{\kappa_o}{4\kappa J}\right)+\mathcal{O}\left(\frac{1}{J^2}\right) \label{eq:ck1}\\
    c_{k+m,1}=\langle E_k| f_{k+m}^{(0)}\rangle&=&\frac{1}{\sqrt{2}}\left(1-\frac{\kappa_o}{4\kappa J}\right)+\mathcal{O}\left(\frac{1}{J^2}\right).\label{eq:ckm1} 
\end{eqnarray}
and 
\begin{eqnarray}
    |\langle E_k| f_{k}^{(0)}\rangle|^2&=&\frac{1}{2}+\frac{\kappa_o}{4\kappa J}+\mathcal{O}\left(\frac{1}{J^2}\right) \label{eq:ck1s}\\
    |\langle E_k| f_{k+m}^{(0)}\rangle|^2&=&\frac{1}{2}-\frac{\kappa_o}{4\kappa J}+\mathcal{O}\left(\frac{1}{J^2}\right). \label{eq:ckm1s}
\end{eqnarray}

\subsection{Dependence on $J$ of  $\langle f_{a'}^{(0)}|f_{a}^{(1)}\rangle$ }

The second and third terms in Eq. (\ref{eq:Fmaxexp}) depend on $\langle f_{a'}^{(0)}|f_{a}^{(1)}\rangle$ given in Eq.(\ref{eq:fofu}).  For $a=k$ and $a'=k+m$, this overlap is
\begin{equation}  \langle f_{k+m}^{(0)}|f_{k}^{(1)}\rangle=-\frac{i}{\phi_{k}^{(1)}-\phi_{k+m}^{(1)}}\left(\sum_{k'\not= k,k+m}\!\!\!\!\! \frac{\langle f_{k+m}^{(0)}|\hat{K}|E_{k'} \rangle\langle E_{k'}|\hat{K}| f_{k}^{(0)}\rangle}{e^{i\tau(E_{k'}-E_{k})}-1} +\frac{1}{2}\langle f_{k+m}^{(0)}|\hat{K}^2|f_{k}^{(0)}\rangle\right).
\label{eq:overAp}
\end{equation}

The first expression inside the parentheses of the previous equation is a sum that, again, is dominated by the terms with small denominators.  Among those terms, the largest ones, for an exact  resonance $m{:}n$, are those with $k'=k-m$ and $k'=k+2m$. For these terms, $\tau(E_{k-m}-E_k)=-2\pi n+2\delta_{m:n}/J$ and $\tau (E_{k+2 m}-E_k)= 4 \pi n+2\delta_{m:n}/J$, which imply $e^{i\tau (E_{k-m}-E_k)}\approx e^{i\tau(E_{k+2m}-E_k)}\approx e^{2i\delta_{m:n}/J}\approx 1+2i\delta_{m:m}/J$.
Therefore, the sum is approximated as
\begin{equation}
\sum_{k'\not= k,k+m}\!\!\!\!\! \frac{\langle f_{k+m}^{(0)}|\hat{K}|E_{k'} \rangle\langle E_{k'}|\hat{K}| f_{k}^{(0)}\rangle}{e^{i\tau(E_{k'}-E_{k})}-1}\approx \frac{\langle f_{k+m}^{(0)}|\hat{K}|E_{k-m} \rangle\langle E_{k-m}|\hat{K}| f_{k}^{(0)}\rangle}{2i\delta_{m:n}}J+\frac{\langle f_{k+m}^{(0)}|\hat{K}|E_{k+2m} \rangle\langle E_{k+2m}|\hat{K}| f_{k}^{(0)}\rangle}{2i\delta_{m:n}}J.
\label{eq:sumExp}
\end{equation}
Now, by using  the  zero order approximations to the Floquet eigenstates, Eqs.\eqref{eq:fapA} and \eqref{eq:fapB}, we obtain that the products in the numerators of the  previous expression are 
$$
\langle f_{k+m}^{(0)}|\hat{K}|E_{k'}\rangle\langle E_{k'}|\hat{K}|f_{k}^{(0)}\rangle\approx \frac{1}{2}\left[|\langle E_{k}|\hat{K}|E_{k'}\rangle|^2-|\langle E_{k+m}|\hat{K}|E_{k'}\rangle|^2 +\frac{\kappa_o}{\kappa J} \langle E_{k+m}|\hat{K}|E_{k'}\rangle\langle E_{k'}|\hat{K}|E_k\rangle\right],
$$
where we have assumed (as it is for the  case considered here) that $\langle E_{k+m}|\hat{K}|E_{k'}\rangle$ and $\langle E_{k'}|\hat{K}|E_k\rangle$ are real numbers.
Substituting this expression in \eqref{eq:sumExp}, we obtain
\begin{align*}
&\sum_{k'\not= k,k+m}\!\!\!\!\! \frac{\langle f_{k+m}^{(0)}|\hat{K}|E_{k'} \rangle\langle E_{k'}|\hat{K}| f_{k}^{(0)}\rangle}{e^{i\tau(E_{k'}-E_{k})}-1}\approx \\
&\frac{|\langle E_{k}|\hat{K}|E_{k-m}\rangle|^2-|\langle E_{k+m}|\hat{K}|E_{k-m}\rangle|^2  ]}{4i\delta_{m:n}}J+
\frac{|\langle E_{k}|\hat{K}|E_{k+2m}\rangle|^2-|\langle E_{k+m}|\hat{K}|E_{k+2m}\rangle|^2  }{4i\delta_{m:n}}J \\
&+\frac{\kappa_o}{4i\delta_{m:n}\kappa  } \left(\langle E_{k+m}|\hat{K}|E_{k-m}\rangle\langle E_{k-m}|\hat{K}|E_k\rangle +\langle E_{k+m}|\hat{K}|E_{k+2m}\rangle\langle E_{k+2m}|\hat{K}|E_k\rangle \right)\\
&\approx
\frac{|\langle E_{k}|\hat{K}|E_{k-m}\rangle|^2 -|\langle E_{k+m}|\hat{K}|E_{k+2m}\rangle|^2 }{4i\delta_{m:n}}J\\
& +\frac{\kappa_o}{4i\delta_{m:n}\kappa  } \left(\langle E_{k+m}|\hat{K}|E_{k-m}\rangle\langle E_{k-m}|\hat{K}|E_k\rangle +\langle E_{k+m}|\hat{K}|E_{k+2m}\rangle\langle E_{k+2m}|\hat{K}|E_k\rangle \right),
\end{align*}
    where in the last step we have neglected matrix elements  between states with  an index distance of $2m$ and kept only those with  an index distance of $m$, $|\langle E_{k}|\hat{K}|E_{k-m}\rangle|\gg |\langle E_{k+m}|\hat{K}|E_{k-m}\rangle|$ and $|\langle E_{k+m}|\hat{K}|E_{k+2m}\rangle|\gg|\langle E_{k}|\hat{K}|E_{k+2 m}\rangle|$. The numerator in the first term of the resulting  expression  depends  on two matrix elements of the same order  that scale quadratically with $J$, but their difference scales only linearly. Together with the factor $J$ coming from the small denominator, we get that this term scales quadratically. The same happens for the second term which depends on the products of two matrix elements.  Accordingly, the whole sum scales quadratically with $J$, that is
$$\sum_{k'\not= k,k+m}\!\!\!\!\! \frac{\langle f_{a'}^{(0)}|\hat{K}|E_{k'} \rangle\langle E_{k'}|\hat{K}| f_{a}^{(0)}\rangle}{e^{i\tau(E_{k'}-E_{a})}-1}= -i B_1  J^2.
$$
Now, we focus on the second expression between the parentheses of Eq. \eqref{eq:overAp} 
$$
\langle f_{k+m}^{(0)}|\hat{K}^2|f_{k}^{(0)}\rangle=\frac{1}{2}\left[ \langle E_k|\hat{K}^2|E_k\rangle-\langle E_{k+m}|\hat{K}^2|E_{k+m}\rangle+\frac{\kappa_o}{\kappa J}\langle E_k|\hat{K}^2| E_{k+m}\rangle\right]
$$
where we have used the zero order Floquet eigenstates at leading order in $1/J$, Eqs.\eqref{eq:fapA} and \eqref{eq:fapB}, and assumed that $\langle E_{k}|\hat{K}^2|E_{k+m}\rangle$ is real. The resulting expression  depends on the difference of  diagonal matrix elements of $\hat{K}^2$. Every matrix element scales quadratically with $J$, but their difference does it only linearly. The last term is divided by $J$, then it only scales linearly with $J$. Consequently, the whole term scales linearly with $J$
$$
\langle f_{a'}^{(0)}|\hat{K}^2|f_{a}^{(0)}\rangle=\frac{1}{2}\left[ \langle E_a|\hat{K}^2|E_a\rangle-\langle E_{a'}|\hat{K}^2|E_{a'}\rangle+\frac{\kappa_o}{\kappa J}\langle E_a|\hat{K}^2| E_{a'}\rangle\right]= B_o J.
$$

Now, according to Eqs.\eqref{eq:quada} and \eqref{eq:quadap}, 
the difference of first order corrections to the  quasienergies  is
$$
\phi_k^{(1)}-\phi_{k+m}^{(1)}\approx -2\kappa J. 
$$
Putting all these results together, we conclude that
\begin{equation}
    \langle f_{a'}^{(0)}|f_{a}^{(1)}\rangle\approx \frac{B_1}{2\kappa}J+ i\frac{B_o}{4\kappa}.
    \label{eq:apLi}
\end{equation}
Based on this scaling behavior of $\langle f_{a'}^{(0)}|f_{a}^{(1)}\rangle$, the terms containing this inner product in Eq.\eqref{eq:Fmaxexp} can be shown to be negligible with respect to the first term in the same equation. This justifies the validity of Eq.\eqref{eq:fmaxapp}, as discussed in the following subsection.

\subsection{Scaling of $\epsilon_\text{max}$ for exact resonances
}

\begin{figure}
    \centering
        \includegraphics[width=0.8\linewidth]{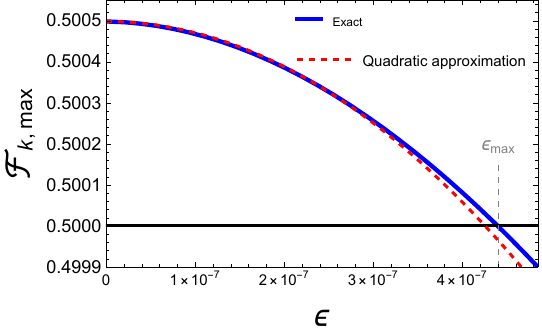}
    \caption{Blue continuous line shows the exact numerical evaluation of    $\mathcal{F}_{k,\text{max}}$ as a function of $\epsilon$ for the exact resonance $1\!:\!1$,  $J=1000$ and $\tau\approx 8$.  Red dashed line is the perturbative approximation up to second order in $\epsilon$, shown in  Eq.~(\ref{eq:fmaxapp}). The condition $\mathcal{F}_{a,\text{max}}=1/2$ (horizontal black line) determines $\epsilon_\text{max}$ (vertical dashed line).}
    \label{fig:fmaxe}
\end{figure}

By substituting the results of the previous subsections [Eqs.~\eqref{eq:SumScaling},\eqref{eq:ck1}-
\eqref{eq:ckm1s} \eqref{eq:apLi}] in Eq.\eqref{eq:Fmaxexp}, we conclude that, for large enough $J$, the dependence on $J$ of  the  maximum fidelity is
$$
\mathcal{F}_{k,\text{max}}=(1-D_o J^4\epsilon^2)\left(\frac{1}{2}+\frac{\kappa_o}{4 \kappa J}\right)+ \frac{B_1}{2\kappa}J\epsilon- \frac{\kappa_o}{2\kappa}\left[\left(\frac{B_1^2}{4\kappa^4}\right)J +\frac{B_o^2}{16\kappa^2 J}\right]\epsilon^2
+ T_2\epsilon^2,$$
where $T_2= 2 \text{Re}[\langle E_a|f_{a}^{(0)}\rangle \langle E_a|f_{a}^{(2)}\rangle^*]$ remains to be determined.
First and  third terms are both quadratic in $\epsilon$, but, since the third term scales linearly with $J$ and the first one scales  as $J^4$, the third term is clearly negligible compared  to the first term for large $J$. On the other hand, the second term  depends on the combination $z=J\epsilon$, whereas the first one depends on $J^4\epsilon^2 =z^2 J^2 $. Therefore, for large enough  $J$, the first term dominates over the second one. We conclude that 
\begin{equation}
\mathcal{F}_{k,\text{max}}=(1-D_o J^4\epsilon^2)\left(\frac{1}{2}+\frac{\kappa_o}{4 \kappa J}\right)+ T_2\epsilon^2.
\label{eq:Fmaxstep}
\end{equation}
The coefficient  $T_2$ requires calculating the second order correction to the Floquet eigenstates in degenerate perturbation theory, a direct but cumbersome task. Instead of that, we  evaluate numerically the relevance of this term by plotting the exact numerical evaluation  of $\mathcal{F}_{k,\text{max}}$ and compare it with the result obtained from the first term in Eq.\eqref{eq:Fmaxstep}. This is done in Fig.~\ref{fig:fmaxe} for an exact resonance $1{:}1$ with $J=1000$. We see that the exact numerical evaluation of $\mathcal{F}_{k,\text{max}}$ is very well described by  the first term in Eq.\eqref{eq:Fmaxstep}, which indicates that the    effect of the term $T_2\epsilon^2$ is rather marginal.  
Therefore
\begin{eqnarray}
\mathcal{F}_{k,\text{max}}\approx(1-D_o J^4\epsilon^2)\left(\frac{1}{2}+\frac{\kappa_o}{4 \kappa J}\right)&=& \left(1-\epsilon^2\sum_{k'\not=a,a'} |\langle E_{k'}|f_a^{(1)}\rangle|^2\right)|\langle E_a|f_{a}^{(0)}\rangle|^2\label{eqq:FmaxAPP}\\
&= &\left(1-\frac{\epsilon^2
}{4}\sum_{k'\not=a,a'}\frac{|\langle E_{k'}|\hat{K}|f_{a}^{(0)}\rangle|^2}{\sin^2\left(\frac{\tau(E_{k'}-E_a)}{2}\right)}\right)|\langle E_a|f_{a}^{(0)}\rangle|^2, \nonumber 
\end{eqnarray}
which is the expression shown in main text with $c=\frac{\kappa_o}{4 \kappa}$ and $D_o=a_{2,\text{ER}}$.

The scaling of $\epsilon_{k,\text{max}}$ for an exact resonant state can be obtained easily  from Eq.\eqref{eqq:FmaxAPP}. 
The maximum perturbation strength is obtained from the condition $\mathcal{F}_{k,\text{max}}=1/2$, which yields  
\begin{equation}    \epsilon_{k,\text{max}}\approx\sqrt{\frac{\kappa_o}{2\kappa J D_o J^4}} = \sqrt{\frac{\kappa_o}{2 \kappa D_o}} \frac{1}{J^{5/2}}=C J^{-5/2}.
\end{equation}
This power-law for $\epsilon_{k,\text{max}}$ is confirmed numerically in Fig.3(c) of main text.

We end by noting that,  in contrast to non-degenerate unitary perturbation theory, where the perturbative series of  $\mathcal{F}_{k,\text{max}}$ up to $\order{\epsilon^2}$ fails to describe the numerical results in the whole range $\epsilon\in[0,\epsilon_{k,\text{max}}]$, degenerate perturbation theory up to $\order{\epsilon^2}$ provides a much more accurate description of $\mathcal{F}_{k,\text{max}}$ in the $\epsilon$-interval of interest. As illustrated in Fig.~\ref{fig:fmaxe}, the second order approximation to $\mathcal{F}_{k,\text{max}}$ in the degenerate case closely matches the numerical results. This comes from the fact that, in the degenerate case,  $\mathcal{F}_{k,\max}$ for infinitesimal $\epsilon$ is larger but very close to $1/2$.      

%% file: LetterWO-GS_v3.bbl
%